\documentclass{article}

\usepackage[version=3]{mhchem} 
\usepackage{siunitx}
\usepackage{amsmath}
\usepackage{hyperref}
\usepackage[capitalize]{cleveref}
\usepackage{multirow}
\usepackage{subfigure}
\usepackage{verbatim}
\usepackage{xr}
\usepackage{authblk}
\usepackage{graphicx}
\usepackage[style=numeric, sorting=none]{biblatex}
\usepackage{fullpage}
\usepackage{caption} 
\DeclareSIUnit\angstrom{\text {Å}}

\bibliography{refs}

\title{Assessing exchange-correlation functionals for heterogeneous catalysis of nitrogen species}
\author[1,2]{Honghui Kim}
\author[2]{Neung-Kyung Yu}
\author[2]{Nianhan Tian}
\author[2]{Andrew J. Medford\thanks{ajm@gatech.edu}}

\affil[1]{Department of Chemical and Biomolecular Engineering (BK21 four), Korea Advanced Institute of Science and Technology (KAIST), Daejeon, Republic of Korea}
\affil[2]{School of Chemical \& Biomolecular Engineering, Georgia Institute of Technology, Atlanta, GA, USA}

\begin{document}

\date{\today}

\maketitle

\begin{abstract}
Increasing interest in sustainable synthesis of ammonia, nitrates, and urea has led to an increase in studies of catalytic conversion between nitrogen-containing compounds using heterogeneous catalysts. 
Density functional theory (DFT) is commonly employed to obtain molecular-scale insight into these reactions, but there have been relatively few assessments of the exchange-correlation functionals that are best suited for heterogeneous catalysis of nitrogen compounds. 
Here, we assess a range of functionals ranging from the generalized gradient approximation (GGA) to the random phase approximation (RPA)  for the formation energies of gas-phase nitrogen species, the lattice constants of representative solids from several common classes of catalysts (metals, oxides, and metal-organic frameworks (MOFs)), and the adsorption energies of a range of nitrogen-containing intermediates on these materials. 
The results reveal that the choice of exchange-correlation functional and van der Waals correction can have a surprisingly large effect and that increasing the level of theory does not always improve the accuracy for nitrogen-containing compounds. 
This suggests that the selection of functionals should be carefully evaluated on the basis of the specific reaction and material being studied.
\end{abstract}

\section{\label{sec:intro}Introduction}
Nitrogen is an essential component of all living things. 
Amino acids and nucleotides, the building blocks of protein and DNA, need nitrogen. 
Air is composed mostly of molecular dinitrogen, but the triple bond between two nitrogen atoms makes it inaccessible to use directly. 
Thankfully, certain bacteria can convert nitrogen gas molecules into useful compounds like ammonia through a process called nitrogen fixation. 
The converted nitrogen is used and transferred to other environments and organisms and then released as gas molecules at the end by the process called denitrification. 
Through these complicated processes, the nitrogen cycle is maintained and is critical for all living things.
Human societies also require fixed nitrogen, particularly in the form of ammonia, for use as fertilizer. 
The Haber-Bosch process meets this demand and has significantly increased the world population. 
However, this has led to a significant artificial increase in fixed nitrogen, disrupting the natural cycle\cite{townsend2003human, vitousek1997human, battye2017nitrogen}. 
Excess nitrogen fixation leads to the generation of greenhouse gases, acid rain, depletion of the ozone layer, and water pollution from nitrite ions \cite{galloway2013chronology, ghaly2015nitrogen, snyder2009review}. 
Thus, the management of the nitrogen cycle has been identified as a scientific ``grand challenge''\cite{zhang2020global, fowler2013global, galloway2008transformation, erisman2013consequences}. 
Significant research has been devoted to overcoming this challenge in recent years, and heterogeneous catalysis, including electrocatalysis and photocatalysis, is a key strategy in efficiently converting nitrogen-containing compounds \cite{zeng2020restoring, xu2022electrocatalytic, soumare2020exploiting, hao2021emerging, li2020vacancy}.

Computational studies are valuable for understanding and predicting heterogeneous catalytic materials.
In the realm of nitrogen chemistry, numerous computational investigations have shed light on specific catalytic materials to synthesize and convert a range of nitrogen compounds \cite{dahl1999role, Jacobsen2001,vojvodic2014exploring, mcenaney2017ammonia, medford2014assessing, logadottir2003ammonia, boisen2005optimal, paolucci2017dynamic,mehta2018overcoming, paolucci2016catalysis, getman2010dft, getman2007dft}. 
Ammonia synthesis is one of the most researched reactions in those studies. 
In computational studies, density functional theory (DFT) has been used to calculate the reaction rate on active catalysts\cite{Honkala2005, Hellman2006}, examine the uncertainty and reliability of computational results\cite{medford2014assessing, kepp2018accuracy}, and propose new catalyst design strategies \cite{Montoya2015, vojvodic2014exploring,Logadottir2001} in ammonia synthesis. 
These successful applications of DFT to accelerate catalyst discovery and provide fundamental insight are not limited to ammonia synthesis. 
Computational techniques have been widely applied to study catalysts for other nitrogen-based reactions such as nitrate reduction\cite{liu2019activity, Mou2022}, urea synthesis\cite{Kong2023, Hanson2021}, and selective catalytic reduction \cite{paolucci2017dynamic, paolucci2014isolation, rostamikia2019elementary, katsounaros2017structure, lim2021structure, wang2021increasing, skulason2012theoretical}, among others.

Discovering high-performance catalysts within the expansive realm of materials poses a formidable challenge. 
The volcano plot, rooted in the Sabatier principle \cite{sabatier1911hydrogenations}, proves invaluable in identifying promising catalyst candidates by representing catalyst activity through adsorption energy descriptors\cite{medford2015sabatier}. 
Taking ammonia synthesis as an illustration, a volcano plot can be constructed using the nitrogen adsorption energy as the descriptor \cite{Logadottir2001, Dahl2001}. 
Researchers use this approach to find materials with optimal \ce{N} adsorption energies to identify candidates with high catalytic activity and cost-effectiveness \cite{singh2018computational, yang2022theoretical}. 
As studies on ammonia synthesis have expanded, the scope of materials of interest (e.g., metal, metal oxide, nitride) and the range of reaction types (e.g., electrocatalysis and photocatalysis) have significantly increased. 
Exploring more intricate materials and reactions requires a mechanistic understanding at the atomic and electronic levels, a task achievable through computational analysis. 
An illustrative example is photocatalytic nitrogen fixation on \ce{TiO2}(110)\cite{Comer2018, comer2018role, huang2023formation}, where a combination of computational and experimental studies revealed the possibility of oxidative or carbon-assisted pathways, expanding the realm of possible mechanisms proposed from purely experimental studies. 
Intensive research is also underway on metal organic frameworks (MOFs) as promising materials for photocatalytic nitrogen fixation, with amine-functionalized MIL-125 serving as a notable example\cite{mohamed2021comprehensive, mohamed2022search}. 
Computational studies of nitrogen reactions on MOFs and other complex materials may yield valuable insights into the relevant mechanisms and strategies for improving catalyst performance.

Given the increasing importance of computational studies in nitrogen-containing compounds, ensuring the accuracy of computationally predicted properties is of utmost importance. 
DFT has become a prevalent tool in such studies and has been employed to compute the physicochemical properties of catalysts. 
While DFT provides computational efficiency compared to higher-level quantum calculations, it exhibits deviations in accuracy. 
This discrepancy arises from the inherent approximation within the exchange-correlation functional, a crucial element for describing electron interactions, leading to divergent computational results with different functionals. 
``Jacob's ladder'' of density functional approximation provides a general perspective on the cost-accuracy tradeoff of functionals\cite{perdew2005prescription}.
Those on the lower rungs (e.g., local density approximation, generalized gradient approximation (GGA)) are relatively fast (with cost typically scaling cubically with the number of atoms), but have larger errors on average. 
Functionals on the higher rungs (e.g., hybrids, random phase approximation (RPA)) are more accurate but can be drastically more computationally expansive due to quadratic or quintic scaling with the number of atoms, practically limiting their routine application to surface science systems.
Nevertheless, the accuracy improvement only reflects a general trend, rather than a systematic guarantee, so the thoughtful selection of a functional tailored to a specific application is essential to achieve precise results with DFT. 
To address this, numerous studies have undertaken benchmarking of exchange-correlation functionals.
Wellendorff {\em et al.} contributed significantly by releasing a benchmark database for adsorption energies on metal surfaces with six functionals \cite{wellendorff2015benchmark} and conducting a comprehensive benchmark on various gas and solid phase properties, comparing the performance of BEEF-vdW with other functionals \cite{wellendorff2012density}. 
Kepp conducted a specific benchmark for ammonia synthesis by modeling a single iron atom as a catalyst \cite{kepp2018accuracy}. 
Araujo {\em et al.} enhanced DFT predictions with corrections from small metal cluster calculations, improving accuracy in periodic systems, and benchmarked it on adsorption energies and activation barriers \cite{araujo2022adsorption}.
Although there are various other benchmark studies\cite{zhao2023benchmark, sorescu2014assessing}, it is important to recognize significant limitations when extrapolating their results to reactions involving nitrogen-containing compounds. 
One of the most critical constraints is the scope of materials and properties investigated in these benchmark studies. 
Metals dominate the benchmarked materials, with only a few exceptions. 
Other physicochemical properties have been benchmarked for metal oxides, but there are relatively few evaluations of chemisorption on oxides, particularly for nitrogen-containing species \cite{castelli2012computational, rasmussen2015computational, gautam2018evaluating, christensen2016functional}.
Furthermore, a notable deficiency in most benchmarks is the absence of an adsorption energy comparison. 
Adsorption energy represents a fundamental property for the calculation of reaction descriptors, the illustration of reaction diagrams, and the prediction of reaction rates.
However, it is notoriously challenging to benchmark due to the challenges in treating adsorption systems with high levels of theory or obtaining highly precise experimental measurements of chemisorption energies, particularly on complex materials \cite{wellendorff2015benchmark}.

In this work, we focus on selected nitrogen reactions, species, and solid catalyst materials relevant for nitrogen chemistry. 
We evaluated gas phase species relevant to a variety of nitrogen reactions (ammonia synthesis, nitrate reduction, urea synthesis, acetamide, nitromethane, and acetonitrile synthesis), and related adsorbed species (\ce{NO}*, \ce{CN}*, \ce{N}*, \ce{N2H}*, \ce{N2}*, \ce{NH2}*, and \ce{NH3}*). 
These adsorbates are selected on the basis of their various types of covalent bonds (\ce{C-N}, \ce{N-O}, \ce{N-H}, \ce{N-N}) and their mechanistic importance in several previous studies of ammonia synthesis and nitrate reduction \cite{abghoui2016electroreduction, zhao2017single, Honkala2005, comer2018role, lim2021structure, liu2019activity}. 
For solid catalysts, we select two examples of metals (\ce{Pd}, \ce{Cu}), metal oxides (\ce{TiO2}, \ce{MoO3}) and MOFs (\ce{MIL-125}, \ce{OCUPUY}). 
These materials are selected based on prior work on nitrate reduction and synthesis of ammonia or urea \cite{perez2017electrocatalytic, lim2021structure, sun2023carbon, comer2018role, Comer2018, Li2019, huang2020toward}, existence of experimental adsorption energies \cite{wellendorff2015benchmark}, and presence of diverse bonding types including van der Waals layers (\ce{MoO3}) and open metal sites (\ce{OCUPUY}). 
For functionals, we select several examples at GGA (PBE\cite{perdew1996generalized}, RPBE\cite{hammer1999improved}), GGA+vdW (BEEF-vdW\cite{wellendorff2012density}, rev-vdW-DF2\cite{hamada2014van}), meta-GGA (mGGA)(SCAN)\cite{sun2015strongly}, and hybrid (B3LYP\cite{stephens1994ab}, PBE0\cite{perdew1996rationale}, HSE06\cite{krukau2006influence}) levels of theory. 
We also utilize RPA optimized to gas-phase atomization energies to provide a ground truth for adsorption energies. 
These functionals are chosen on the basis of their common use in the literature and the diversity of physically derived and empirically fitted functionals. 
For all functionals without explicit vdW corrections, we also evaluate the influence of the empirical D3 parameters for dispersion forces. 
For gas-phase species, we compare to experimental formation energies, and for solid-state materials, we compare to experimental lattice constants. 
For adsorption energies, we compare to an optimized RPA result that, based on gas-phase performance, is expected to be close to chemical accuracy.
Although the approach is far from exhaustive, it provides representative diversity in the electronic structures that are typical in nitrogen catalysis and the functionals commonly used to treat them, yielding the most systematic view of the role of exchange-correlation functionals on nitrogen catalysis to date.

The results are roughly consistent with the conventional wisdom that functional choice leads to variations of $\sim$ 0.2-0.4\SI{}{eV} in adsorption energies, with a standard deviation of \SI{0.355}{eV} across all functionals, materials, and adsorbates (or \SI{0.243}{eV} if the OCUPUY MOF with significant functional-dependent geometric differences is omitted). However, the errors compared to optimized RPA results are considerably larger (RMSE of \SI{0.548}{eV}). 
The findings are also consistent with the expectation that the B3LYP hybrid functional has strong performance for gas phase formation energies (RMSE of \SI{0.086}{eV}) and reaction energies (RMSE of \SI{0.100}{eV})\cite{mardirossian2017thirty}.  
Interestingly, the B3LYP functional also yields excellent results for lattice constants, despite previous reports that hybrid functionals perform poorly for metals \cite{paier2007does, furche2006performance}. 
We also find that the standard RPA functional yields results worse than B3LYP (RMSE of \SI{0.110}{eV}), but that a few simple optimizations enable chemical accuracy (RMSE of \SI{0.035}{eV}) on gas-phase formation energies with RPA correlation. 
Moreover, we find a surprising influence of dispersion corrections on adsorption energies, even with small molecules, leading to some counterintuitive conclusions. 
For example, as expected, the RPBE functional is one of the weaker binding functionals on average \cite{wellendorff2015benchmark}. 
However, when D3 corrections are included, it becomes the strongest binding functional, with D3 corrections that can exceed \SI{1}{eV}, even for small adsorbates. 
We also find that, although the standard deviation between functionals is $\sim$ 0.2-0.4\SI{}{eV}, specific deviations in the adsorption energies of individual species can vary by $>$ \SI{1}{eV} with different functionals. 
These deviations are not systematic across different nitrogen reactions, making it important to carefully select a functional based on the reaction of interest. 

\section{\label{sec:methods}Methods}

\subsection{\label{sec:methods_DFT}Details of electronic structure calculations}

Density Functional Theory (DFT) calculations were performed using the Vienna {\em ab initio} Simulation Package (VASP) v6.1.2, employing projector-augmented wave (PAW) pseudopotentials (VASP v5.4.4 was used for BEEF-vdW calculations)\cite{kresse1993ab, kresse1996efficiency, kresse1996efficient, kresse1999ultrasoft}. 
Dispersion corrections were incorporated in all calculations using the DFT-D3 method with the Becke-Johnson damping function\cite{Grimme2010, Grimme2011}, except for calculations that employ rev-vdW-DF2, BEEF-vdW or RPA, which inherently account for dispersion. 
The pre-defined damping function parameters for each exchange-correlation functional in VASP were used, with Grimme et al.'s parameters applied for HSE06\cite{Moellmann2014} (see \Cref{tab: d3_param}).
We omit +U corrections from all functionals due to ambiguities in the implementation of the method and the meaning of the U parameters across different implementations \cite{santana2014successes, wang2016local, rosen2022high}.

A k-point spacing of \SI{0.4}{\per\angstrom} and a kinetic energy cutoff of \SI{600}{\eV} were used for bulk and slab models of \ce{Cu} and \ce{Pd}. 
For \ce{TiO2}, \ce{MoO3}, \ce{OCUPUY}, and \ce{MIL-125}, a k-points spacing of \SI{0.5}{\per\angstrom} and a kinetic energy cutoff of \SI{600}{\eV} were used. 
The $\Gamma$ point was included in all k-point grids. 
A kinetic energy cutoff of \SI{600}{\eV} was used to calculate the molecular formation energy and the molecular reaction energy. 
Convergence tests for the kinetic energy and k-points are provided in \Cref{tab:convergence_setup}, \Cref{tab:convergence_setup_kpts}, \Cref{tab:convergence_test_solids}, and \Cref{tab:convergence_test_gases}.
Gaussian smearing of band occupancies with a smearing width of \SI{0.05}{\eV} was used.
The convergence criterion of \SI{e-5}{\eV} in total energy was applied to the self-consistent field (SCF) cycle. 
All calculations were spin-polarized. 
Non-spherical contributions related to the gradient of the density in the PAW spheres were included for hybrid functionals, SCAN, rev-vdW-DF2, and BEEF-vdW. 
The periodic and atomic cell relaxations continued until the forces on each atom were less than \SI{0.03}{\eV/\angstrom}.

Initially, each unit cell of bulk materials was relaxed to determine the optimized lattice constants and atomic positions. 
We used the initial structures of the two MOFs\cite{dan2009new, Ouellette2006} from the CoRE MOF 2019 database \cite{chung2019advances}.
However, MIL-125 in CoREMOF 2019 database has deprotonated oxygen atoms, which is unphysical. 
We protonated the deprotonated MIL-125 as reported in prior papers\cite{walsh2010photostimulated, hendon2013engineering}. 
In the case of MOFs, the most stable spin state configuration was determined prior to lattice relaxation, with the tested spin states listed in \Cref{tab:optimal_spin_config_OCUPUY}. 
Using optimized cell and atomic positions, slab models for \ce{Cu} (100 facet), \ce{Pd} (111 facet), \ce{TiO2} (110 facet), and \ce{MoO3} (100 facet) were constructed in four layers, with the bottom two layers kept fixed in all subsequent calculations. 
To avoid self-interaction between repeated slabs, \SI{10}{\angstrom} of vacuum is added to the z-direction.
For the slab models and the two MOFs, a gas molecule was introduced, and all atomic positions were optimized while the lattice constants was kept fixed. 
The initial placement of gas molecules was determined by manual placement on the basis of prior literature reports and intuition. 
The geometry optimization of adsorbates was performed solely for five non-hybrid functionals, and for three hybrid functionals and RPA functionals, single-point calculations were performed using PBE-optimized geometry. 
To obtain the energy of a molecule in DFT, a gas molecule was placed in the center of a cubic box with a \SI{10}{\angstrom} lattice and its geometry was optimized using a $\Gamma$ point calculation.

All RPA calculations were performed using the low-scaling RPA algorithm \cite{kaltak2014cubic} implemented in VASP v6.3.2. 
The GW pseudopotentials were used with 24 frequency points. 
For nonmetals, GW (H) or GW\_new (C, N, O) pseudopotentials were used, and for metals, sv\_GW pseudopotentials were used except for Cu, which was calculated using a GW pseudopotential with lower valency. 
The errors arising from pseudopotentials are expected to be $<$ \SI{0.05}{eV} on average (see \Cref{fig:rpa_conv_cu}). For the calculation of DFT orbitals used in the evaluation of RPA correlation energy, the total energy was converged to below \SI{e-8}{\eV} during the SCF cycle.

In a standard RPA calculation, the total energy is given by the equation, 
\begin{equation}
    E_{total} = E_{HF}+E_{c, RPA}
\end{equation}
where $E_{HF}$ and $E_{c, RPA}$ refer to the Hartree-Fock energy and the RPA correlation energy, respectively, both evaluated non-self-consistently using DFT orbitals. The choice of DFT functionals leads to different orbitals and different RPA energies. RPA\text{@}PBE, RPA\text{@}PBE0, and RPA\text{@}PBEx50 refer to RPA calculations using different DFT orbitals obtained with PBE, PBE0, and a tuned hybrid functional, respectively. 
The tuned hybrid functional, PBEx50, is based on PBE0, but the amount of Fock exchange is set to 50 \% as opposed to the 25 \% used in PBE0. Previous reports have shown that including additional Fock exchange can improve the results of GW calculations\cite{kaplan2016gw}.
We also use an optimized version of RPA using the results of RPA with PBEx50 orbitals, where the RPA correlation energy is multiplied by a simple linear scaling factor of 1.17 (i.e. $E_{optRPA} = E_{HF}+1.17E_{c, RPA}$) .
The scaling factor for $E_{c,RPA}$ was determined to minimize errors in molecular atomization energies, as shown in \Cref{fig:rpa_atomization}, based on the observation that the absolute contributions of $E_{c,RPA}$ to both the molecular atomization energy and the formation energy are underestimated in RPA calculations. It has previously been reported that the magnitude of $E_{c,RPA}$ can be incorrect \cite{hong2007EcRPA}, and applying the scaling factor to $E_{c,RPA}$ leads to highly accurate results for molecular energies. In the main text, we refer to this optimized version of RPA as optRPA, and it is always based on the PBEx50 orbitals.

For molecules, all RPA calculations were performed with a single $\Gamma$ point, and molecules were placed inside cubic cells (with lattice lengths of \SI{15}{\angstrom} for $E_{HF}$ and \SI{10}{\angstrom} for $E_{c,RPA}$). 
Two different approaches were employed to calculate the gas formation/reaction energy and adsorption energy. 
To obtain the gas formation/reaction energy, the RPA energy was evaluated at the PBE optimized structures, and different kinetic energy cutoffs (\SI{700}{\eV} for $E_{HF}$ and \SI{600}{\eV} for $E_{c,RPA}$) were used to obtain the two energy terms. 
For the adsorption energy, the relevant molecules were calculated with the same energy cutoffs as those used for the surfaces to maximize cancellation of error, and we verified that these lower cutoffs led to average numerical errors of $<$0.02 eV in molecular formation energies.

The energy of surface systems was determined at the PBE-optimized structures, with kinetic energy cutoffs of \SI{600}{\eV} for $E_{HF}$ and \SI{400}{\eV} for $E_{c,RPA}$, based on the energy convergence test (see \Cref{fig:rpa_conv_ecut}). 
In the case of RPA calculations, varying formalisms, energy cutoffs, and smearing were used to achieve an expected numerical accuracy of \SI{0.05}{\eV} per system. 
Details of convergence tests for gas-phase, nonmetallic, and metallic systems are provided in the SI.

\subsection{Evaluation}

The evaluation metrics for the benchmarking comprised both gas- and solid-phase properties. 
The properties of the gas-phase molecules included molecular formation energy and molecular reaction energy. 
Molecular formation energy was computed using a least-squares regression approach (see \cref{equ:linreg} and \cref{equ:formation}) to minimize systematic errors.
Least-squares regression aligns calculated total energies of molecules ($E_i$) as closely as possible to experimental formation energy values ($\Delta_f E_i$) using atomic stoichiometries ($n_{i,j}$) and chemical potentials ($\mu_j$) to minimize residual errors ($\epsilon_i$). This approach ensures that all physically observable properties (i.e. energies of mass-balanced reactions) are consistent with the underlying method, but minimizes non-systematic cancellation of error. The resulting least-squares formation energies are denoted as $\Delta_f^{LS}E_i$. 
Reaction energies were calculated using \cref{equ:rxn}.
All experimental energies of gas molecules were taken from the \SI{0}{K} ATcT \cite{ruscic2005active}, and zero-point corrected using frequencies from the NIST Webbook\cite{linstorm1998nist} and CCCBDB\cite{cccbdb}.

\begin{equation}
    \begin{aligned}
        \label{equ:linreg}
        &\begin{pmatrix}
        \Delta_f E_1^{exp} \\
        \Delta_f E_2^{exp} \\
        \vdots \\
        \Delta_f E_n^{exp}
        \end{pmatrix}=
        \begin{pmatrix}
        E_1  & n_{1,C} & n_{1,H} & n_{1,O} & n_{1,N} \\
        E_2 & n_{2,C} & n_{2,H} & n_{2,O} & n_{2,N} \\
        \vdots & \vdots & \vdots & \vdots & \vdots \\
        E_n & n_{n,C} & n_{n,H} & n_{n,O} & n_{n,N}
        \end{pmatrix}
        \begin{pmatrix}
        1 \\
        -\mu_C \\
        -\mu_H \\
        -\mu_O \\
        -\mu_N
        \end{pmatrix}
        +\begin{pmatrix}
        \epsilon_1 \\
        \epsilon_2 \\
        \vdots \\
        \epsilon_n
        \end{pmatrix}\\
    \end{aligned}
\end{equation}

\begin{equation}
    \label{equ:formation}
    \begin{aligned}
    \Delta_f E_i^{exp} &= E_i-n_{i,C}\mu_C-n_{i,H}\mu_H-n_{i,O}\mu_O-n_{i,N}\mu_N +\epsilon_i \\
    &= \Delta^{LS}_f E_{i} +\epsilon_i 
    \end{aligned}
\end{equation}

\begin{equation}
\label{equ:rxn}
    \begin{aligned}
        \Delta_r E &= \sum_{products}n_i\Delta_f E_{i} - \sum_{reactants}n_j\Delta_f E_{j}\\&=\sum_{products}n_iE_i - \sum_{reactants}n_jE_j
    \end{aligned}
\end{equation}

For the solid-phase and interfacial properties, we computed the unit cell volume and gas adsorption energies.
The volume of each DFT-optimized unit cell is compared to that of the experimental unit cell by normalizing the volume per atom.
The experimental volume is preprocessed to directly compare it with the DFT-computed volume, by subtracting the thermal expansion volume and zero-point effect.
Due to lack of experimental data, this thermal correction was not possible for MOFs, so the reported room-temperature lattice constants were used.
The detailed methods and data used to correct the volume are in SI. The formation energy of the adsorbed species was calculated using \cref{equ:ads}, anchored to the reference species \ce{N2}, \ce{H2}, \ce{H2O}, and \ce{CH4}.
\begin{equation}
    \label{equ:ads}
    \Delta_f^{'}E_{i,ads} = E_{solid+ads}-E_{solid}-E_{ref}
\end{equation}
where $E_{ref}$ is a gas-phase reference based on \ce{N2}, \ce{H2}, \ce{H2O}, and \ce{CH4}. 

Standard deviation evaluates the consistency between functionals in each crystal+gas system, except for RPAs.
Total standard deviation ($\sigma_{total}$) of each material is computed by using \cref{equ:stddev}.
\begin{equation}
    \label{equ:stddev}
    \sigma_{total} = \sqrt{\frac{\Sigma\sigma_{system}^2}{\text{number of systems}}}
\end{equation}

\section{\label{sec:results}Results}

\subsection{\label{sec:molecularform}Molecular Formation Energy}

\begin{figure}[ht]
    \centering
    \includegraphics[width=\textwidth]{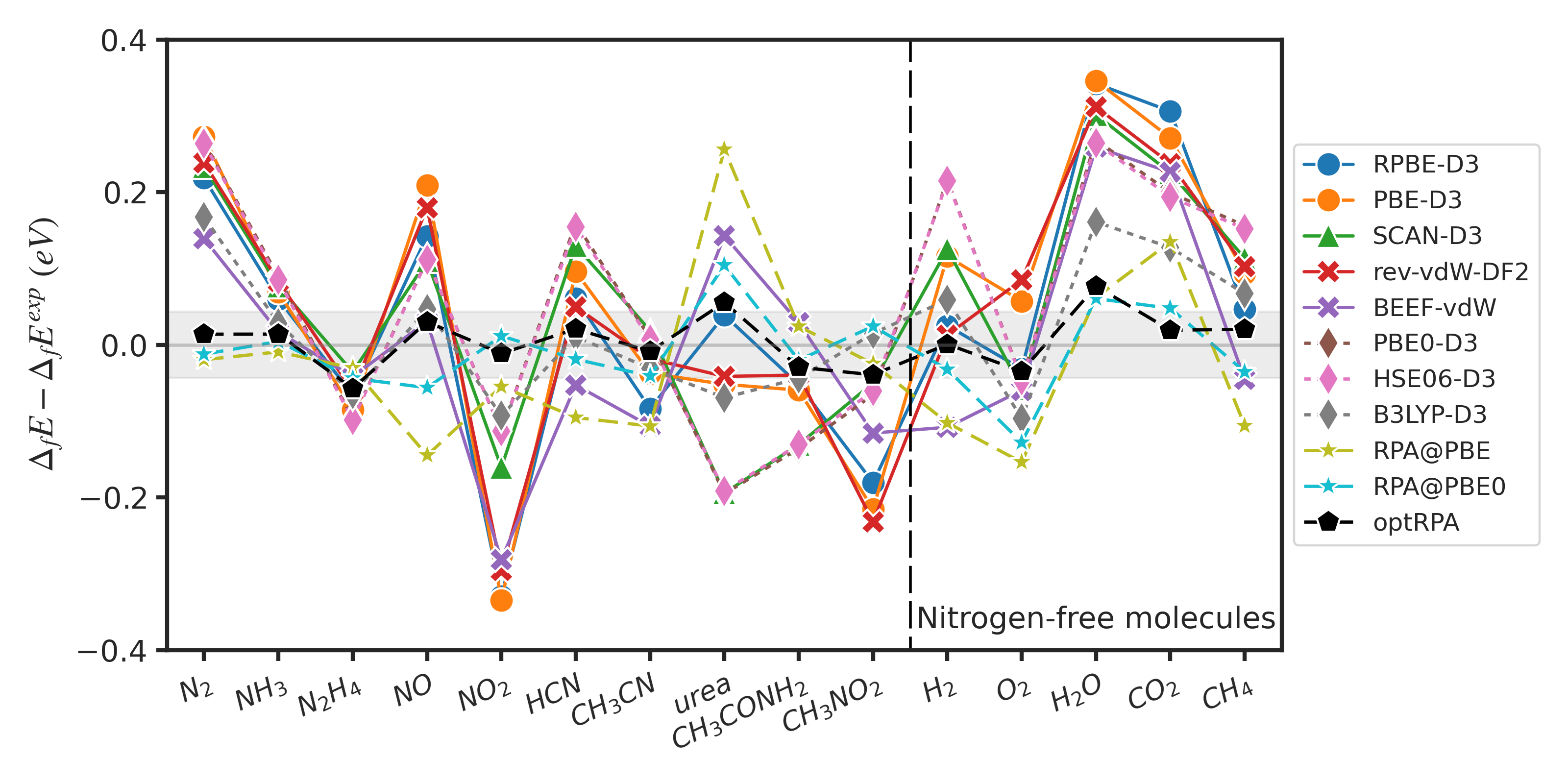}
    \caption{Error of the calculated formation energies from the experimental formation energies. All formation energies are computed by linear regression to eliminate systematic errors. The grey shaded region represents chemical accuracy, ±0.0433 eV (±1 kcal/mol). PBE0-D3 plot is eclipsed by HSE06-D3 plot.}
    \label{fig:formation}
\end{figure}

We selected ten nitrogen-containing species (\ce{CH3NO2}, \ce{NO2}, \ce{NO}, \ce{CH3CN}, \ce{NH3}, \ce{N2}, \ce{HCN}, \ce{CH3CONH2}, \ce{N2H4}, \ce{NH2CONH2}) and five other species (\ce{CO2}, \ce{H2}, \ce{O2}, \ce{H2O}, \ce{CH4}) commonly found in nitrogen-containing reactions (e.g., water splitting, \ce{CO2} reduction). 
\Cref{fig:formation} and \Cref{tab:formation_reaction_err} illustrate the deviation of calculated formation energy from the experimental formation energy. 
The optRPA functional has the smallest error in all metrics, and the RMSE of optRPA (\SI{0.035}{eV}) is within chemical accuracy (\SI{0.043}{eV}), and has a maximum error magnitude of \SI{0.077}{eV}. 
Interestingly, the standard RPA (RPA@PBE) results are not significantly better than hybrid functionals (RMSE: \SI{0.110}{eV}), and B3LYP-D3 exhibits the highest accuracy (RMSE: \SI{0.086}{eV}) except for optRPA and RPA@PBE0 (RMSE: \SI{0.054}{eV}), consistent with its well-known strong performance for gas-phase molecules\cite{mardirossian2017thirty}. 
The other hybrid functionals, PBE0-D3 and HSE06-D3, yield very similar results to SCAN-D3, and actually exhibit slightly higher errors than SCAN-D3. 
At the GGA level of theory, BEEF-vdW yields the lowest errors, outperforming even SCAN-D3 and some hybrids, but notably, all 15 gas species were included in the training set for BEEF-vdW. 
The rev-vdW-DF2 and RPBE-D3 results are similar and slightly outperform PBE-D3, which yields the largest errors for gas-phase formation energies. 

\begin{table}[ht]\centering
\setlength{\tabcolsep}{4pt}
\renewcommand{\arraystretch}{1.3}
  \caption{Error of the calculated formation energies and reaction energies compared to the experimental formation and reaction energies. Unit in \SI{}{eV}.}
  \label{tab:formation_reaction_err}
  \begin{tabular}{l|*{3}{c}|*{3}{c}}
    \hline
    \hline
    \multirow{2}{*}{}   & \multicolumn{3}{c|}{$\Delta^{LS}_f E - \Delta_f E^{exp}$} & \multicolumn{3}{c}{$\Delta_r E - \Delta_r E^{exp}$}\\
    \cline{2-7}
                        & MaxError & MAE& RMSE & MaxError & MAE& RMSE   \\ 
    \hline
            RPBE-D3     &  $0.342$ & $0.132$ &  $0.173$ & $-0.456$ &  $0.194$&  $0.248$\\
            PBE-D3      &  $0.346$ & $0.154$ &  $0.187$ & $-0.572$ &  $0.210$&  $0.273$\\
            SCAN-D3     &  $0.301$ & $0.131$ &  $0.152$ & $-0.272$ &  $0.169$&  $0.195$\\
            rev-vdW-DF2 &  $0.312$ & $0.133$ &  $0.167$ & $-0.518$ &  $0.194$&  $0.254$\\
            BEEF-vdW    &  $-0.282$ & $0.110$ &  $0.138$ & $0.515$ &  $0.242$&  $0.299$\\
            PBE0-D3     &  $0.268$ & $0.142$ &  $0.160$ & $-0.368$ &  $0.207$&  $0.245$\\
            HSE06-D3    &  $0.264$ & $0.139$ &  $0.157$ & $-0.369$ &  $0.206$&  $0.243$\\
            B3LYP-D3    &  $0.167$ & $0.072$ &  $0.086$ & $0.150$ &  $0.082$&  $0.100$\\
            RPA@PBE     &  $0.256$ & $0.088$ &  $0.110$ & $0.453$ &  $0.213$&  $0.245$\\
            RPA@PBE0    &  $-0.128$ & $0.043$ &  $0.054$ & $0.202$ &  $0.112$&  $0.128$\\
            optRPA      &  $0.077$ & $0.029$ &  $0.035$ & $0.094$ &  $0.051$&  $0.060$\\
    \hline
    \hline
  \end{tabular}
\end{table}

\subsection{\label{sec:molecularreac}Molecular Reaction Energy}

\begin{figure}[ht]
    \centering
    \includegraphics[width=\textwidth]{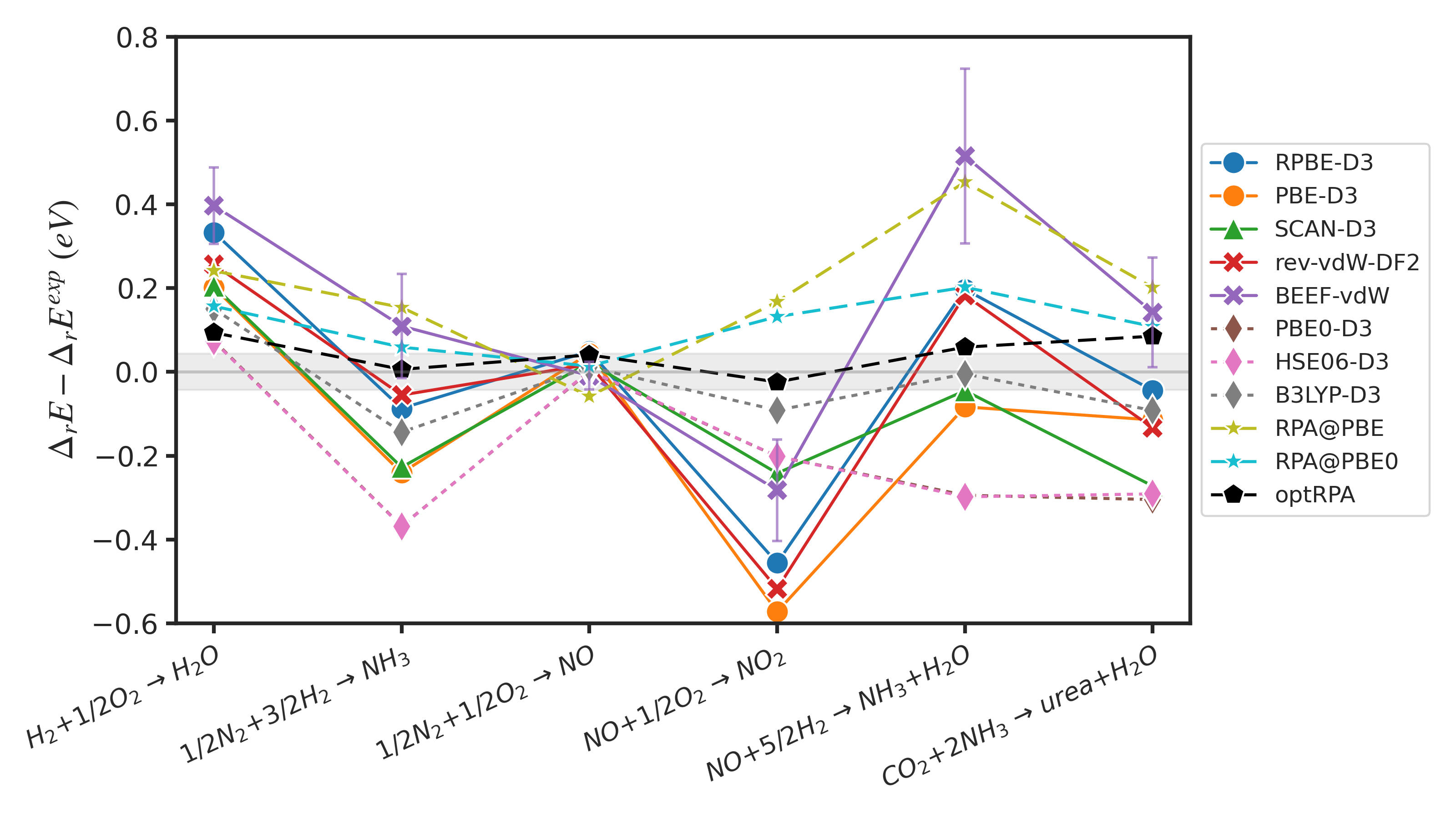}
    \caption{Error of the calculated reaction energies from the experimental reaction energies. Grey shade is a region of the chemical accuracy, \SI{\pm0.043}{eV} (\SI{\pm1}{kcal/mol}). PBE0-D3 plot is eclipsed by HSE06-D3 plot.}
    \label{fig:reaction}
\end{figure}

Reaction energies provide a practical metric for the accuracy of gas-phase properties, as they are directly relevant to chemical reactions. 
They differ from formation energies in that cancellation of error occurs implicitly in the reaction energy calculation and can thus yield different results from formation energy benchmarks. 
Here, we select six common reactions in nitrogen chemistry. 
The difference between the calculated and experimental reaction energies for these reactions is depicted in \Cref{fig:reaction}, and the corresponding errors for each functional are presented in \Cref{tab:formation_reaction_err}. 
Trends are broadly consistent with the formation energy results, with optRPA and B3LYP-D3 emerging as the most accurate overall. 
Interestingly, B3LYP-D3 (RMSE: \SI{0.100}{eV}) outperforms RPA@PBE0 (RMSE: \SI{0.128}{eV}) in the reaction energy, mainly due to large error of RPA@PBE0 for \ce{NO} reduction to \ce{NH3}.
optRPA exhibits errors slightly higher than chemical accuracy on average (RMSE: \SI{0.060}{eV}), demonstrating a peerless accuracy compared to others.
The results of standard RPA@PBE are far worse, with maximum and average errors comparable to GGA functionals. 
Additionally, contrary to the formation energy results, BEEF-vdW shows the largest errors, despite the fact that the species involved were included in the training set. 
The BEEF-vdW error is largely driven by the reaction for \ce{NO} reduction to \ce{NH3},  which was not included in BEEF-vdW training. 
In these reactions, the importance of individual species differs from the reactions used in training, and the poor performance of BEEF-vdW in these cases illustrates how (lack of) error cancellation can strongly influence the performance of exchange-correlation functionals \cite{stevanovic2012correcting, wang2006oxidation, medvedev2017density}. 
The accuracy of SCAN-D3 also improves for reaction energies, significantly outperforming even PBE0-D3 and HSE06-D3, indicating strong error cancellation for SCAN-D3. 
All other GGA functionals exhibit similar performance, with rev-vdw-DF2 and RPBE-D3 being slightly more accurate than PBE-D3. 
Notably, these conclusions may differ if other reactions are selected, but the focus here is on evaluating common nitrogen chemistry reactions.

One advantage of the BEEF-vdW functional is its ability to provide error estimates. 
These error estimates reflect the sensitivity to the GGA exchange enhancement factor and are generally expected to reflect variations that arise from selecting different GGA functionals. 
However, it is clear that this is not always the case, with RPBE-D3 and PBE-D3 often falling well outside the error bars of BEEF-vdW. 
This may be due to effects of different geometries between the functionals, the fact that BEEF-vdW error estimates are not based on self-consistent densities, or the inclusion of D3 corrections in other GGA functionals. 
Other classes of functionals, particularly hybrids and RPA functionals, are often even further outside the error estimates of BEEF-vdW \cite{szaro2023benchmarking}, indicating that BEEF-vdW error bars should be interpreted with caution. 
Nevertheless, there is a general correlation between the BEEF-vdW error estimate and the standard deviation of all functionals for a given reaction, suggesting that, although the error estimates are not quantitatively accurate, they reliably capture the relative sensitivity to functional choice. 

Another interesting observation from the reaction energies is the high accuracy of reaction energies for NO synthesis from \ce{N2} and \ce{O2} for all functionals. 
This is remarkable, given the fact that \ce{O2} has a triplet spin state and that \ce{NO} contains an unpaired electron. 
These complex electronic structures are generally difficult to treat with DFT, and \ce{O2} is commonly ``corrected'' by 0.4 - 0.5 eV to ensure the correct energetics of the water splitting reaction \cite{norskov2004origin}. 
The high accuracy of \ce{NO} synthesis with all functionals, along with the relatively large errors for other reactions involving \ce{NO} and \ce{O2}, suggests that there is reliable error cancellation between \ce{O2} and \ce{NO}, and reveal that care should be taken if any \ce{O2} correction is applied in cases where \ce{NO} synthesis also occurs.

\subsection{\label{sec:lattice}Lattice Constants}

\begin{figure}[ht]
    \centering
    \includegraphics[width=\textwidth]{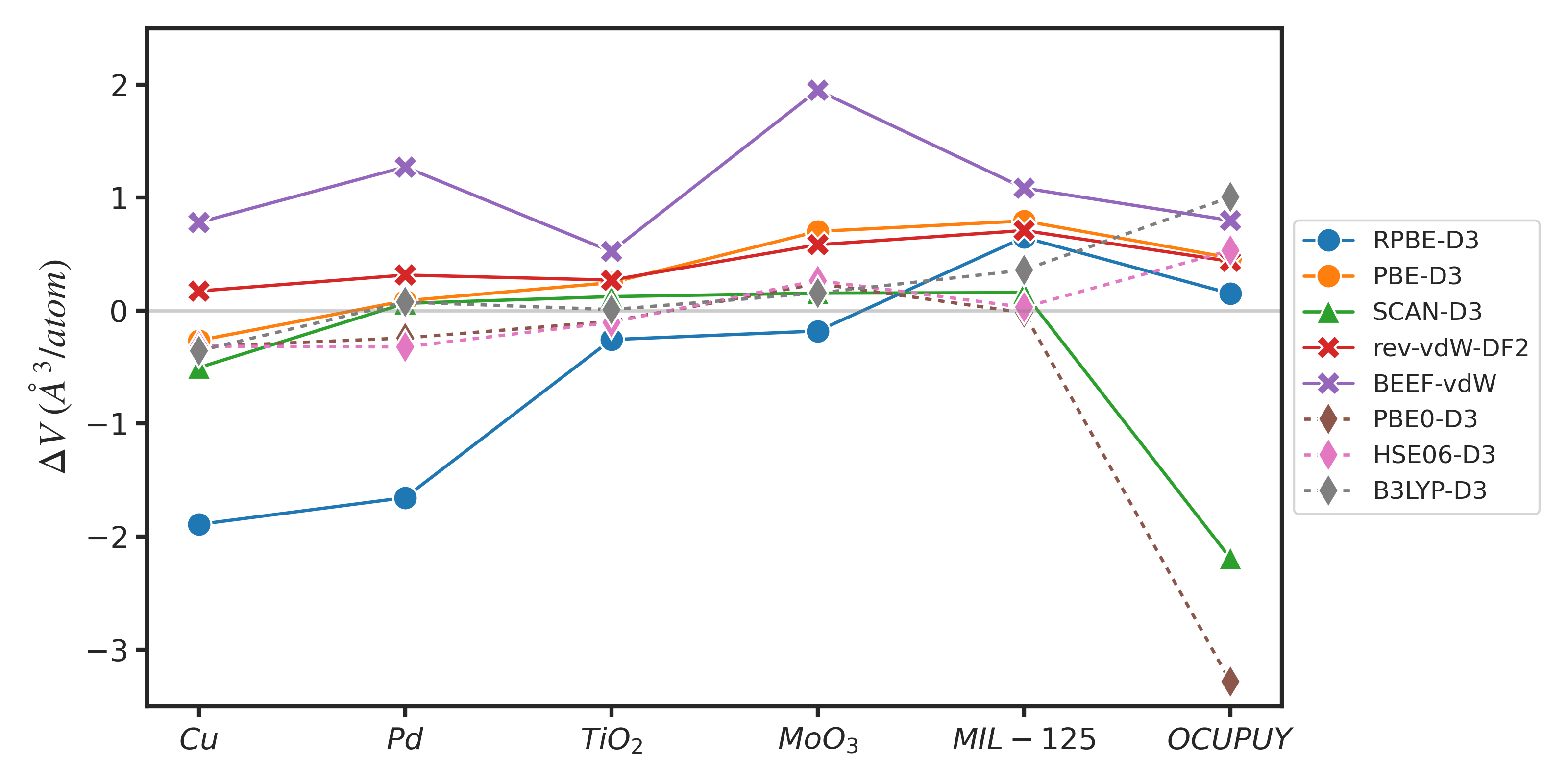}
  \caption{Error of the calculated unitcell volume per atom from the experimental unitcell volume.}
  \label{fig:volume}
\end{figure}

Unit cell volume is selected as a metric for assessing solid-state performance of functionals, because experimental values are generally available for all materials classes of interest. 
A comparison between the calculated and experimental unit cell volume per atom and errors is shown in \Cref{fig:volume} and \Cref{tab:volume}. 
The OCUPUY MOF exhibits considerable variance because of the potential for multiple geometries of the water molecules in the structure. 
This leads to inconsistent optimization of the unit cell and atom positions, with SCAN-D3 and PBE0-D3 predicting different geometries of water than other functionals. 
These effects are difficult to deconvolute, as discussed in the Discussion, so OCUPUY was excluded from the computation of max error, MAE, and RMSE to provide a more fair comparison of performance. 
Across the five materials, B3LYP-D3 shows the smallest RMSE, even for metallic systems, contrary to some previous reports on the applicability of B3LYP to solid and metallic systems\cite{paier2007does, furche2006performance}. 
The HSE06-D3, PBE0-D3, and SCAN-D3 functionals also have comparable accuracy to B3LYP-D3. 
Of the GGA functionals, rev-vdW-DF2 and PBE-D3 demonstrate relatively high accuracy, while BEEF-vdW and RPBE-D3 have substantially larger errors. 
The low accuracy of BEEF-vdW may be due to the fact that oxides and MOFs were not included in the training set, although the error on metals is also higher than that of most other functionals. 
The poor performance of RPBE-D3 is attributed to the D3 corrections, as discussed further in the Discussion. 
We note that the very small number of solid-state materials included here and the fact that only cell volume is evaluated makes it difficult to draw strong conclusions about the specific ranking of functionals for solid-state properties. 
However, the diversity of materials classes evaluated is expected to yield general qualitative insight into the suitability of these functionals for the treatment of solids.

\begin{table}[ht]\centering
\setlength{\tabcolsep}{8pt}
\renewcommand{\arraystretch}{1.3}
  \caption{Max error in magnitude, mean signed error, and root mean squared error of the calculated unitcell volume per atom from the experimental unitcell volume. Numbers in paranthesis are the errors calculated without MIL-125 since thermal corrections are not available for MIL-125. OCUPUY is not included in any error calculations due to inconsistent geometries. Units are \SI{}{\angstrom^3/atom}.}
  \label{tab:volume}
  \begin{tabular}{l|*{3}{c}}
    \hline
    \hline
                        & MaxError & MSE& RMSE   \\ 
    \hline
        RPBE-D3          & $-1.892$ ($-1.892$) & $-0.667$ ($-0.996$) & $1.170$ ($1.267$) \\
        PBE-D3           & $0.794$ ($0.701$) & $0.313$ ($0.193$) & $0.502$ ($0.397$) \\
        SCAN-D3          & $-0.505$ ($-0.505$) & $-0.000$ ($-0.040$) & $0.254$ ($0.273$) \\
        rev-vdW-DF2      & $0.709$ ($0.583$) & $0.410$ ($0.336$) & $0.457$ ($0.368$) \\
        BEEF-vdW         & $1.951$ ($1.951$) & $1.122$ ($1.132$) & $1.224$ ($1.256$) \\
        PBE0-D3          & $-0.320$ ($-0.320$) & $-0.087$ ($-0.105$) & $0.213$ ($0.238$) \\
        HSE06-D3         & $-0.321$ ($-0.321$) & $-0.089$ ($-0.118$) & $0.239$ ($0.266$) \\
        B3LYP-D3         & $0.360$ ($-0.355$) & $0.049$ ($-0.029$) & $0.239$ ($0.198$) \\
    \hline
    \hline
  \end{tabular}
\end{table}

\subsection{\label{sec:adsorption}Adsorption Energies}

\begin{figure}
  \centering
  \begin{subfigure}
    \centering
    \includegraphics[width=\linewidth]{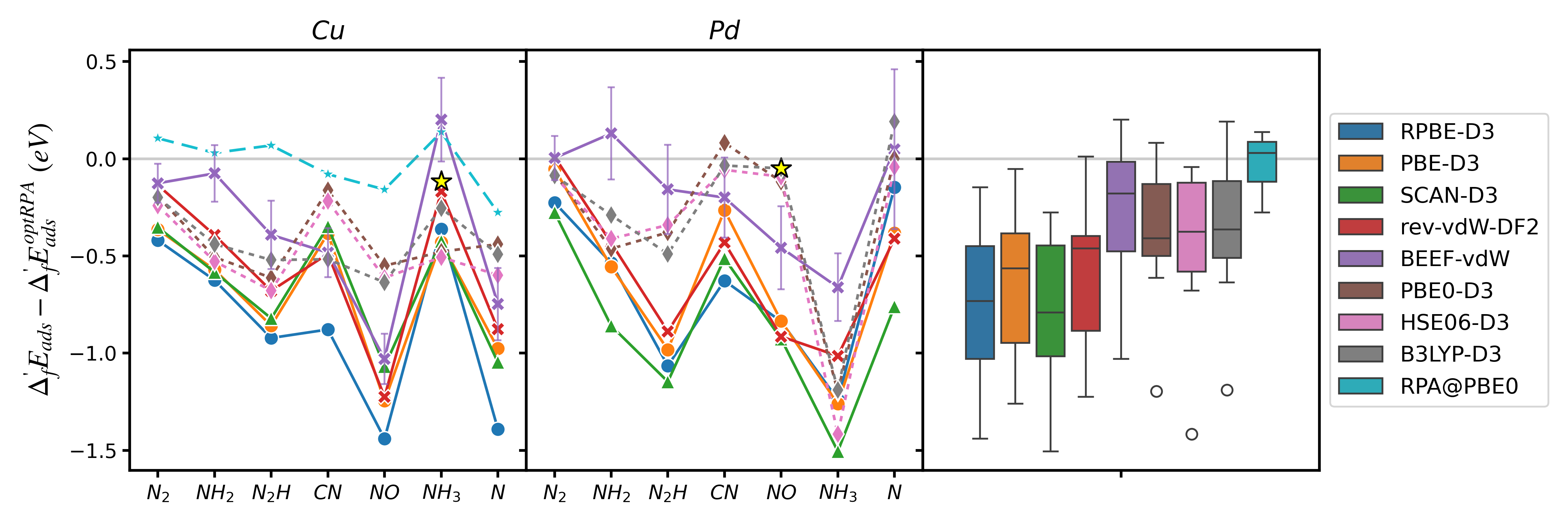}
  \end{subfigure}
  \hfill
  \begin{subfigure}
    \centering
    \includegraphics[width=\linewidth]{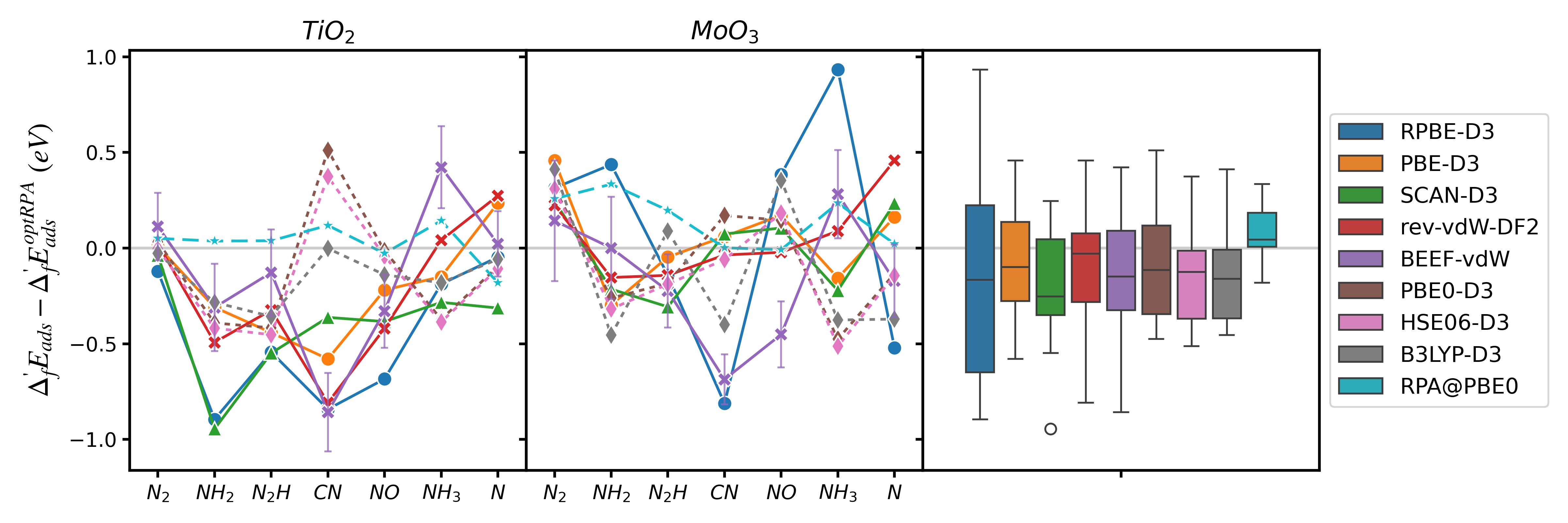}
  \end{subfigure}
  \hfill
  \begin{subfigure}
    \centering
    \includegraphics[width=\linewidth]{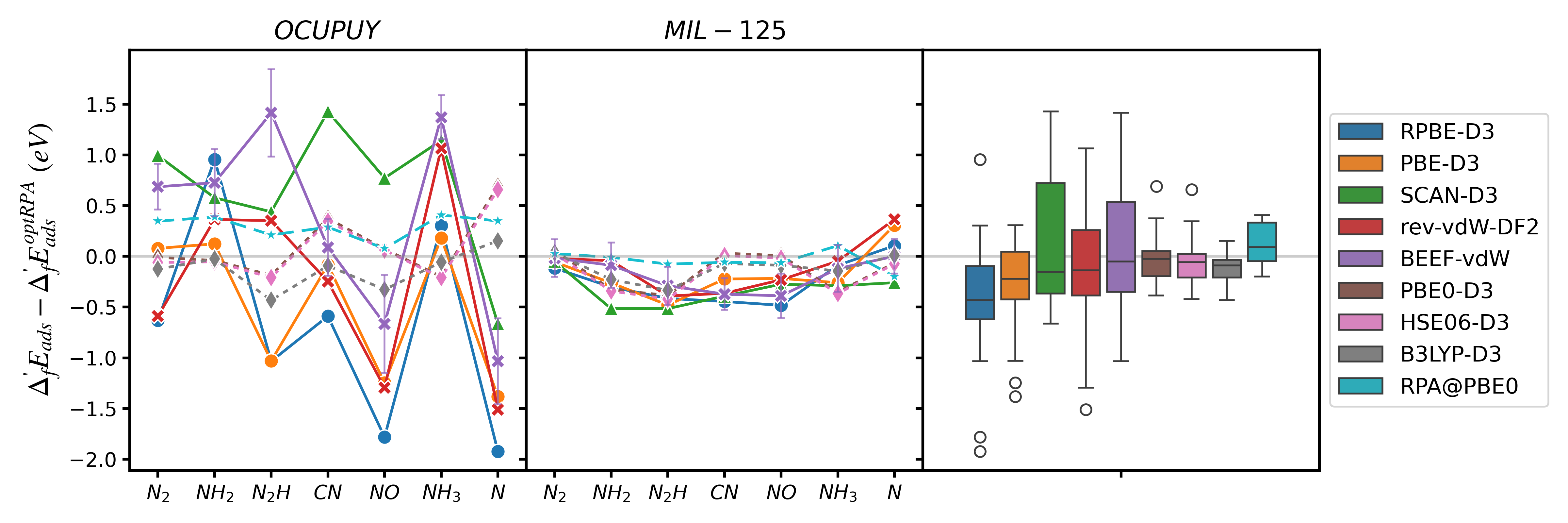}
  \end{subfigure}
  \caption{Error of the calculated adsorption energies in (top) metals, (center) metal oxides, and (bottom) MOFs from the adsorption energy of optRPA. The distribution of errors for each functional are plotted on the right as box and whisker plots. The top, middle, and bottom line of the box indicates 75\% quartile (Q3), median, and 25\% quartile (Q1) respectively. The whiskers extend from the edge of the box to the smallest and largest values within 1.5 times the interquartile range of Q1 and Q3, respectively. All three hybrid functionals and optRPA numbers are calculated by running single-point calculation on PBE-optimized geometry. Yellow stars in the plot indicate the reported experimental adsorption energies corrected to be directly comparable to DFT energies. For \ce{Pd}, the adsorption energy from RPA@PBE is set to be the ground truth because optRPA calculations require significantly higher computational cost due to the use of the densest k-point grids.}
  \label{fig:adsorption}
\end{figure}

Adsorption energies for several nitrogen-containing intermediates (\ce{N}*, \ce{N2}*, \ce{CN}*, \ce{NO}*, \ce{N2H}*, \ce{NH2}*, \ce{NH3}*) calculated at different levels of theory on all six materials are illustrated in \Cref{fig:adsorption}. 
Given the absence of experimental adsorption energy data for most systems, we utilize the optRPA result as the ``ground truth'' on the basis of its excellent performance for gas phase energies (see \Cref{tab:SIads_E_RPAs} for the calculated adsorption energies from optRPA together with other RPA variants).
The commonly used RPA@PBE approach differs from RPA@PBE0 by an RMSE of \SI{0.217}{eV}, and from optRPA by an RMSE of \SI{0.297}{eV} for systems where all RPA results are available (all \ce{Cu} systems and \ce{NH2}, \ce{N2H}, and \ce{CN} on \ce{TiO2}).
The errors of all other calculated adsorption energies compared to optRPA are listed in \Cref{tab:adsorption_err} and \Cref{tab:adsorption_stddev_err}.
Calculated adsorption energies from all four RPAs (optRPA, RPA@PBEx50, RPA@PBE0, RPA@PBE) are listed in \Cref{tab:SIads_E_RPAs}.
For systems with known experimental values (Cu(100)+\ce{NH3}*, Pd(111)+\ce{NO}*), the experimental adsorption energy is marked as a yellow star, and is very close to the optRPA result. 
As discussed in the Methods, all hybrid and RPA calculations are obtained through single-point calculations using PBE-D3 optimized geometry. 
The numbers in this section represent a practical comparison of adsorption energies that would be obtained if standard practices in the literature were followed; a more detailed discussion of the influence of geometries and spin states is provided in the Discussion.

In general, the deviations of adsorption energies between functionals for metals and oxides exhibit standard deviations $\sim$\SI{0.2}{eV}, consistent with the commonly accepted exchange-correlation error of 0.2 - 0.3 eV. 
The standard deviations of all functionals for metals and metal oxides are \SI{0.251}{eV} and \SI{0.274}{eV} respectively.
Interestingly, if optRPA is assumed to be the ground truth, metals have a higher RMSE (\SI{0.654}{eV}) compared to metal oxides (\SI{0.365}{eV}).
All functionals systematically overestimate the adsorption strength in metals with an average mean signed error (MSE) of \SI{-0.522}{eV}.
There is also an overestimation trend in metal oxides, but the amount of overestimation is much smaller (MSE: \SI{-0.134}{eV}).  
On the other hand, RPA@PBE0 tends to bind weakly, with MSEs ranging from \SI{-0.025}{eV} on metals to \SI{0.128}{eV} on MOFs, and exhibits RMSEs of \SI{0.143}{eV} for metals, \SI{0.157}{eV} for oxides, and \SI{0.233}{eV} for MOFs.
Caveats regarding optRPA adsorption energies are discussed in the Discussion. 
For metals, the BEEF-vdW functional has the lowest adsorption energy error (RMSE: \SI{0.447}{eV}) compared to optRPA, which is consistent with the fact that it was trained on metal adsorption energies, although very few of the systems evaluated here were included in the training set \cite{wellendorff2012density}. 
Similar to the case of gas-phase energies, the BEEF-vdW error bars generally do not encapsulate other functionals, although their size loosely correlates with the spread of all functionals with a Pearson correlation coefficient of 0.536 (see \Cref{fig:SIbeef}). 
In metals, hybrid functionals have performance similar to BEEF-vdW and each other, with PBE0-D3 and B3LYP-D3 being slightly more accurate than HSE06-D3. 
On the other hand, all other GGA functionals and SCAN strongly over-bind, with rev-vdW-DF2 having the best performance, and PBE-D3 being slightly worse. 
Surprisingly, RPBE-D3 shows the worst performance and significantly overbinds in all cases, which is true even in metal oxides except for a few instances on \ce{MoO3}. 
This is contrary to the expectation that RPBE binds relatively weakly and is due to D3 corrections, as discussed further in the Discussion. 
Interestingly, the tendency to overbind is stronger on metals than on metal oxides for all functionals, and all functionals are more accurate for oxide adsorption energies than metals, despite the fact that no +U corrections are applied. 
However, for metal oxides the relative performance of functionals shifts, and hybrids are more accurate than BEEF-vdW. 
In \ce{TiO2}, B3LYP-D3 exhibits the lowest error among the eight functionals (RMSE of \SI{0.194}{eV}). 
The PBE-D3, PBE0-D3 and rev-vdW-DF2 functionals exhibit relatively little systematic error on oxides (MSE $<$ \SI{0.1}{eV}), and rev-vdW-DF2 performs particularly well (RMSE of \SI{0.211}{eV}) for the \ce{MoO3} system with vdW layers. 
RMSE for each material is listed in \Cref{tab:SIadsorption_err_material}.

In the case of MOFs, the errors and deviations are large because of OCUPUY, with a total standard deviation between all functionals of \SI{0.489}{eV}, and a RMSE relative to optRPA of \SI{0.583}{eV} across all functionals. 
The situation is also quite different between the two MOFs chosen in this study. 
In the case of OCUPUY, there is relatively little systematic error compared to optRPA (MSE = \SI{-0.041}{eV}), but there is a very large deviation between the functionals (standard deviation of \SI{0.678}{eV}). 
On the other hand, for MIL-125 there is a relatively larger systematic error compared to optRPA (MSE = \SI{-0.183}{eV}), but a smaller deviation between functionals (standard deviation of \SI{0.140}{eV}).
This reflects a key difference in the two MOFs, since adsorption occurs at a spin-polarized open metal site in OCUPUY, but occurs in the pores of MIL-125. 
Furthermore, as discussed in the results of lattice constants, OCUPUY contains several weakly bound water molecules that adopt different geometries for different functionals. 
The effects of geometry and magnetism are discussed in more detail in the Discussion, but here we focus on the deviations that would be observed in a typical high-throughput study with standardized settings across materials. 
In the case of MIL-125, the adsorption energies are much more consistent between functionals, with a small standard deviation of \SI{0.140}{eV} between functionals, and a maximum error of $\sim$\SI{0.5}{eV} when compared to optRPA.
Among non-RPA functionals, B3LYP-D3 has the smallest error (RMSE of \SI{0.169}{eV}).
The hybrid functionals and the non-hybrid functionals have similar errors in MIL-125 while the hybrid functionals are much more accurate in OCUPUY.
This is consistent with previous reports that hybrid functionals required for open metal sites \cite{rosen2022high, grajciar2010water}.
These findings indicate that chemisorption in MOFs with open metal sites may exhibit errors much larger than those of metal or metal oxide systems.
We found that the source of the large error includes not only differences between functionals but also variations in geometry, though small amounts of magnetism also contribute.
The effects of different geometry and magnetism are discussed in the following Discussion section.

\begin{table}[ht]\centering
\setlength{\tabcolsep}{4pt}
\renewcommand{\arraystretch}{1.3}
  \caption{Error of the calculated adsorption energy compared to the adsorption energy of optRPA. Unit in \SI{}{eV}.}
  \label{tab:adsorption_err}
  \begin{tabular}{l|*{3}{c}|*{3}{c}|*{3}{c}}
    \hline
    \hline
    \multirow{2}{*}{}   & \multicolumn{3}{c|}{Metals} & \multicolumn{3}{c|}{Metal Oxides} & \multicolumn{3}{c}{MOFs} \\
                        & MaxError & MSE& RMSE   & MaxError & MSE& RMSE   & MaxError & MSE& RMSE\\ 
    \hline
            RPBE-D3     &  $-1.440$ & $-0.765$ & $0.864 $&  $0.933 $ & $-0.195$ & $0.572 $&  $-1.922$ & $-0.460$ & $0.863 $\\
            PBE-D3      &  $-1.259$ & $-0.654$ & $0.747 $&  $-0.579$ & $-0.079$ & $0.285 $&  $-1.382$ & $-0.325$ & $0.607 $\\
            SCAN-D3     &  $-1.505$ & $-0.759$ & $0.835 $&  $-0.945$ & $-0.211$ & $0.375 $&  $1.428 $ & $0.171 $ & $0.698 $\\
            rev-vdW-DF2 &  $-1.225$ & $-0.575$ & $0.674 $&  $-0.808$ & $-0.092$ & $0.335 $&  $-1.510$ & $-0.185$ & $0.667 $\\
            BEEF-vdW    &  $-1.029$ & $-0.282$ & $0.447 $&  $-0.857$ & $-0.155$ & $0.378 $&  $1.416 $ & $0.094 $ & $0.697 $\\
            PBE0-D3     &  $-1.197$ & $-0.364$ & $0.481 $&  $0.512 $ & $-0.084$ & $0.297 $&  $0.691 $ & $-0.023$ & $0.279 $\\
            HSE06-D3    &  $-1.416$ & $-0.417$ & $0.544 $&  $-0.513$ & $-0.126$ & $0.297 $&  $0.657 $ & $-0.046$ & $0.278 $\\
            B3LYP-D3    &  $-1.190$ & $-0.358$ & $0.483 $&  $-0.455$ & $-0.128$ & $0.294 $&  $-0.432$ & $-0.123$ & $0.197 $\\
            RPA@PBE0    &  $-0.277$ & $-0.025$ & $0.143 $&  $0.334 $ & $0.087 $ & $0.157 $&  $0.408 $ & $0.128 $ & $0.233 $\\
    \hline
    \hline
  \end{tabular}
\end{table}

\begin{table}[ht]\centering
\setlength{\tabcolsep}{10pt}
\renewcommand{\arraystretch}{1.3}
  \caption{Standard deviation and error of the calculated adsorption energy compared to the adsorption energy of optRPA, for each material. ``std. dev'' stands for standard deviation between functionals (except for RPA results). Unit in \SI{}{eV}.}
  \label{tab:adsorption_stddev_err}
  \begin{tabular}{l|*{3}{c}}
    \hline
    \hline
                        & std. dev & MSE& RMSE \\ 
    \hline
            \ce{Cu}         & $0.221$ & $-0.563$ & $0.655$ \\
            \ce{Pd}         & $0.277$ & $-0.481$ & $0.652$ \\
            \ce{TiO2}       & $0.263$ & $-0.237$ & $0.397$ \\
            \ce{MoO3}       & $0.284$ & $-0.031$ & $0.331$ \\
            \ce{OCUPUY}     & $0.678$ & $-0.041$ & $0.780$ \\
            \ce{MIL-125}    & $0.140$ & $-0.183$ & $0.270$ \\
    \hline
            Total           & $0.355$ & $-0.256$ & $0.548$ \\
    \hline
    \hline
  \end{tabular}
\end{table}

\section{\label{sec:discussion}Discussion}

\subsection{\label{sec:vanderwaals}Influence of Van der Waals Corrections}

The impact of the D3 correction on calculated adsorption energies was found to be substantial, showing notable variations between different functionals and systems. 
To examine the contribution of the D3 correction, we compared properties with and without the correction.
The latter were obtained by subtracting the D3 correction from the D3 corrected results.
As expected, molecular formation energies and molecular reaction energies are relatively insensitive to the D3 correction (see \Cref{tab:formation_reaction_err}, \Cref{tab:formation_reaction_err_nod3}, \Cref{fig:SIformation}, and \Cref{fig:SIreaction}), although the D3 correction has a noticeable effect on the gas-phase properties calculated by RPBE-D3. 
This is because the parameters of RPBE ($s_8$, $a_1$, and $a_2$) used in \cref{equ:d3bj2} for the D3 correction make the calculated dispersion energy large. 
Interestingly, almost all functionals have slightly increased RMSE and MAE for gas phase properties when the D3 correction is added.
However, the difference is $\sim$\SI{0.02}{eV} in RMSE, which is comparable to the numerical accuracy of the calculations.

\begin{equation}
E_\text{disp} = -\frac{1}{2} \sum_{i=1}^{N_\text{at}} \sum_{j=1}^{N_\text{at}} \sum_{L}{}' \left( f_\text{d,6}(r_{ij,L}) \frac{C_{6ij}}{r_{ij,L}^6} + f_\text{d,8}(r_{ij,L}) \frac{C_{8ij}}{r_{ij,L}^8} \right)
\label{equ:d3bj}
\end{equation}

\begin{equation}
f_d,n(r_{ij}) = \frac{s_n r_{ij}^n}{r_{ij}^n + \left( a_1 R_{0ij} + a_2\right)^n}
\label{equ:d3bj2}
\end{equation}

In contrast, the optimized lattice constants are significantly impacted by dispersion correction. 
To test the effect of D3 correction on unit cell volume, we performed cell relaxations without the D3 correction on metals and metal oxides using PBE (see \Cref{tab:vol_w_wo_d3}). 
Due to the absence of dispersion interactions, cell relaxation without D3 correction results in expanded volume compared to \Cref{fig:volume}. 
The amount of expansion is significant, so the discrepancy from the experimental volume becomes larger. 
In particular, the lattice constant of \ce{MoO3}, a material with van der Waals layers, changes drastically ($\Delta V =\SI{2.215}{\angstrom^3/atom}$) without D3 correction. 
The change was even greater with RPBE ($\Delta V =\SI{3.288}{\angstrom^3/atom}$) due to the larger magnitude of the parameters used in \cref{equ:d3bj2}. 
The conclusion is that the D3 corrections generally improve the volumes of solids.

\begin{figure}[ht!]
    \centering
    \includegraphics[width=\textwidth]{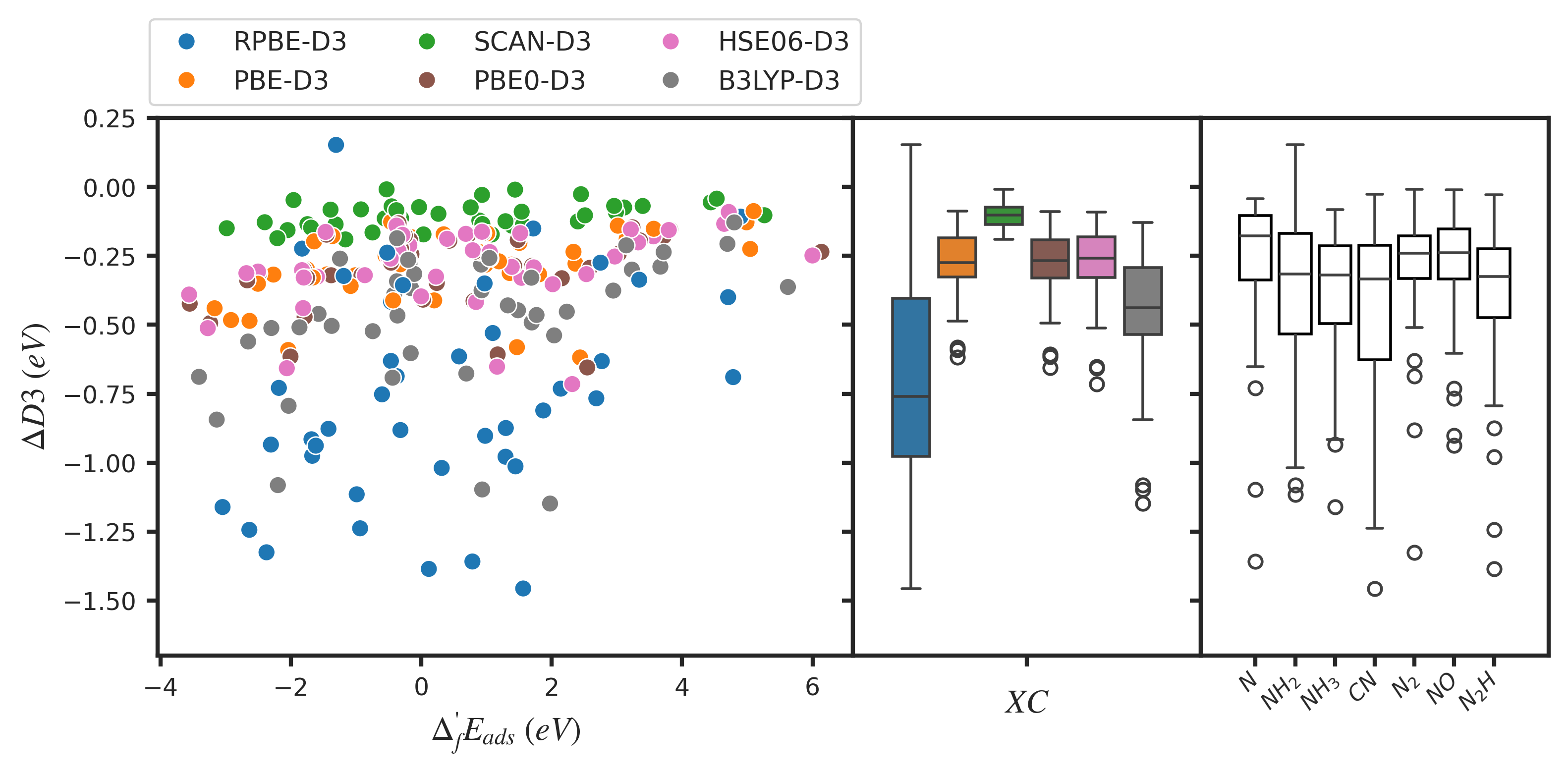}
    \caption{Magnitude of the D3 correction as a function of the adsorption energy for all systems and functionals. Negative sign in the y-axis means that the adsorbate-adsorbent interaction is energetically favorable.}
    \label{fig:vdw}
\end{figure}

Despite relatively small changes in gas phase energies and improvements in solid-state lattice constants, the effects of D3 corrections on adsorption energies are significant, highly varied, and difficult to assess. 
Surprisingly, the change in adsorption energy after adding the D3 correction can exceed \SI{1}{eV} depending on the systems and functionals. 
Because the D3 correction adds attractive dispersion interactions, adsorption becomes stronger when the D3 correction is included in all cases except for \ce{NH2} adsorption on \ce{MoO3} using RPBE-D3. 
This exception occurs due to reconstruction of atoms in \ce{MoO3} after the adsorption of \ce{NH2}. 
\Cref{fig:vdw} shows the distribution of D3 corrections as a function of the adsorption energies, with box plots showing deviation by functional and adsorbate. 
Interestingly, there is only a very weak correlation between the magnitude of the D3 correction and the magnitude of adsorption or the size of the molecule, revealing that contrary to conventional wisdom, the D3 correction can have a significant influence even in cases of strong chemisorption of atoms or small molecules. 
 
There is also a significant variation in the magnitude of D3 corrections between functionals. 
In the case of RPBE, the magnitude and variance of the D3 correction was much larger than other functionals, with an average contribution of \SI{-0.743}{\eV}, causing RPBE-D3 to be one of the strongest binding functionals. 
However, if the D3 correction is removed, then RPBE becomes one of the weakest binding functionals (\Cref{fig:SIadsorption_nod3}), in agreement with prior work\cite{wellendorff2015benchmark}. 
Based on available experimental adsorption energies (\Cref{tab:exp_comparison}) and optRPA results, it appears that the D3 correction overcorrects the adsorption energies for RPBE.
B3LYP has the second largest D3 correction, \SI{-0.470}{\eV} on average, which yields more accurate results for the limited number of experimental adsorption energies.
The D3 corrections for PBE, PBE0 and HSE06 are smaller in magnitude and similar to each other ($\sim$\SI{-0.29}{\eV} on average), while SCAN has the smallest correction (\SI{-0.103}{\eV} on average). 
The corrections for these functionals are more consistent with the typical magnitude of physisorption, and give adsorption energies similar to rev-vdW-DF2, so they are likely more accurately capturing dispersion in the adsorption systems.

Notably, in VASP the D3 correction also influences both energies and forces and can significantly affect the geometry of the system, especially when dispersion interactions are crucial to maintain the crystal structure. 
To deconvolute the effects of geometry and D3 dispersion energy, it is necessary to compare the adsorption energy with the interaction energy, which is the energy difference with all geometric coordinates fixed\cite{sriram2023open}. 
We compare these two quantities for three representative cases: (i) \ce{CN}+\ce{MoO3} with strong chemisorption and dispersion interactions in the solid, (ii) \ce{CN}+\ce{Pd}(111) with strong chemisorption and no significant dispersion forces in the solid and (iii) \ce{N2}+\ce{Pd}(111) with dispersion forces dominating the adsorption of the weakly bound adsorbate (see \Cref{tab:d3_interaction}).
In the case of \ce{MoO3}+\ce{CN}, the geometry changes played a significant role. 
The D3 correction in the interaction energy (\SI{-0.57}{eV}) is much smaller than in the adsorption energy (\SI{-1.46}{eV}), indicating that the dispersion-induced reconstruction of \ce{MoO3} contributes significantly to the D3 correction. 
However, in the cases of \ce{N2} and \ce{CN} adsorption on \ce{Pd} (111), the contribution of D3 is independent of geometry. 
This indicates that the dominant effect in most cases is the energetic contribution of the D3 correction, rather than the geometry changes.

\subsection{\label{sec:geometry}Influence of Geometry and Spin States}

Adsorption energies are the critical parameter in computational studies of heterogeneous catalysis, and are hence the primary quantity evaluated here.
However, while properties such as unit cell volume and gas state energies are straightforward to compute with a single DFT simulation, calculating adsorption energies requires multiple steps and presents a more intricate challenge in determining the accuracy and precision of each functional. 
An adsorption energy calculation involves lattice relaxation to find the optimal unit cell of the solid, computing the energy of the gas phase molecule (or reference), computing the energy and geometry of the adsorbent without the adsorbate present, and finally calculating the energy of the adsorbate-adsorbent system. 
Each of these steps involves an optimization of both the atomic positions and magnetic moments (if the system is magnetic), both of which may differ between functionals and/or between the adsorbent systems with and without the adsorbate present. 
Differences in geometry or magnetic moments between any of the subsystems involved will be included in the adsorption energy. 
These differences in geometry and magnetic moments may reflect real physical differences in the system but may also lead to artifacts where not all systems are in their global minimum state. 
This issue is exacerbated in comparison of functionals, where different functionals may have significantly different geometric and/or magnetic ground states. 
The adsorption energies presented in the Results section encompass all of these effects and are meant to provide practical insight into the deviations in adsorption energies between functionals in high-throughput scenarios where standard input settings are used. 
In this section, we briefly outline scenarios where the indirect geometric and/or magnetic effects may be as or more significant than the direct impact of the energy functional.

Different exchange-correlation functionals always have slight discrepancies in geometric configurations, but in some cases they lead to entirely different geometries that are not equivalent. 
For example, \ce{NO} adsorption on \ce{MoO3} exhibits two distinct local minima, chemisorption and physisorption, where specific GGA functionals (PBE-D3, RPBE-D3) capture both states while others converge only to chemisorption regardless of the initial guess used (see \Cref{tab:SIMoO3_NO_physi_chemi}). 
Additionally, the relaxation of OCUPUY's cell reveals varied geometries for water molecules binding to vanadium atoms depending on the functional used, significantly impacting the optimized volume of OCUPUY. 
Retaining solvent molecules affects the chemical environment of the open metal site, influencing electronic properties and applications such as gas adsorption. 
Often computational studies are conducted on the clean MOFs by removing all solvent molecules perfectly for consistency and simplification, but this can lead to atomic geometries of open metal sites that are not achievable in real experiments.
The process also commonly leads to artifacts such as deprotonated linkers that can drastically influence results. 
Notably, these artifacts are present even for well-established databases such as CoRE MOF 2019 and common MOF materials such as MIL-125. 
\Cref{fig:SIadsorption_deprotonated_MIL125} shows the drastic ($>$ \SI{3}{eV}) differences in adsorption energies that arise when the deprotonated structure from CoRE MOF 2019 is used directly, compared to the properly protonated structure.
Furthermore, the original OCUPUY paper reported water crystallization in the framework \cite{Ouellette2006}, so solvent molecules were retained in OCUPUY in this work. 
Lattice optimization of OCUPUY yields two distinct local minima (\Cref{fig:SIOCUPUY_water}), primarily due to variations in the orientation of the water molecule bound to vanadium open metal sites, leading to significant variance among functional calculations and introducing errors not solely attributed to the accuracy of the exchange-correlation functional. 
Despite significant effort, it was not practically feasible to force OCUPUY to conform to a consistent geometry between functionals during the lattice optimization, highlighting that these geometric effects are often unavoidable in practice.

\begin{figure}[ht!]
    \centering
    \includegraphics[width=\textwidth]{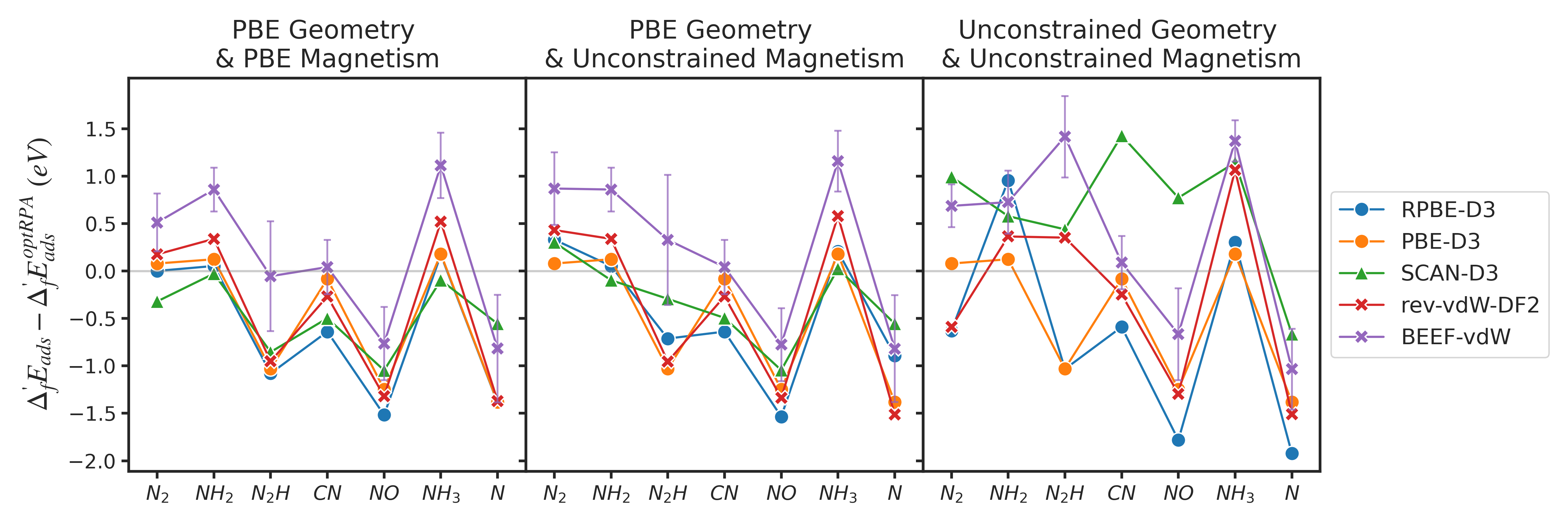}
    \caption{Error of the calculated adsorption energies in OCUPUY from the adsorption energy of optRPA calculation, with the (left) fixed PBE-optimzed geometry and PBE-optimized magnetic moments, (center) the fixed PBE-optimized geometry but relaxed magnetic moments, and (right) all relaxed systems (same with the OCUPUY results in \Cref{fig:adsorption}). PBE Geometry refers to geometry from PBE-D3 optimization.}
    \label{fig:adsorption_OCUPUY_fix}
\end{figure}

\begin{table}[ht]\centering
\setlength{\tabcolsep}{8pt}
\renewcommand{\arraystretch}{1.2}
    \caption{Comparison of discrepancy in calculated adsorption energies between geometry-constrained OCUPUY (fixed PBE geometry) and unconstrained OCUPUY (original), relative to fully constrained OCUPUY (fixed PBE geometry and PBE magnetism). Unit in \SI{}{eV}.}
    \begin{tabular}{l|*{3}{c}}
        \hline
        \hline
           & MaxError & MAE &  RMSE\\
        \hline
        & \multicolumn{3}{c}{fixed PBE geometry}\\
         \hline
        RPBE-D3        &0.497 &0.178&  0.265\\
        SCAN-D3        &0.624 &0.198&  0.321\\
        rev-vdW-DF2    &0.256 &0.068&  0.113\\
        BEEF-vdW       &0.383 &0.114&  0.199\\
         \hline
         &\multicolumn{3}{c}{Original (unconstrained)}\\
         \hline
        RPBE-D3        &0.903 &0.366&  0.477\\
        SCAN-D3        &1.926 &1.187&  1.328\\
        rev-vdW-DF2    &1.304 &0.404&  0.610\\
        BEEF-vdW       &1.470 &0.343&  0.578\\
         \hline
         \hline
    \end{tabular}
    \label{tab:OCUPUY_geomag}
\end{table}

In addition, the transition-metal atom at open metal sites in MOFs often have unpaired electrons, resulting in multiple magnetic states (high-spin or low-spin). 
Within a MOF unit cell, each metal atom possessing unpaired electrons can adopt different magnetic configurations, either up or down, leading to a combinatorial problem to determine optimal spin configuration. 
In high-throughput settings, this optimization is typically performed for the empty MOF structure\cite{sriram2023open}, but magnetic configurations can change during gas molecule adsorption. 
For example, in OCUPUY, we verified that the optimal spin state configuration comprises all up spin states for the vanadium atoms (\Cref{tab:optimal_spin_config_OCUPUY}), yet this optimal state is not always retained during gas adsorption processes (\Cref{tab:OCUPUY_gas_mag}). 
In numerous computational studies, the focus has been on determining the optimal spin state of the MOF adsorbent \cite{lee2015small, nazarian2015benchmarking}. 
However, the majority of high throughput studies have not accounted for potential changes in the optimal spin state when gas molecules are adsorbed.
Therefore, \Cref{fig:adsorption} represents the variations that are likely present in results generated with high-throughput computational workflows.
Similarly to the case of geometry differences, these changes in magnetic moment may differ between functionals, further convoluting the effect of the exchange-correlation functional. 
Notably, Hegde {\em et al.} reported that magnetization is one of the most difficult properties to reproduce across all classes of solid-state materials, even when working with the same input parameters as the database that reports these properties \cite{hegde2023quantifying}. 
This indicates that the adsorption energies for magnetic systems will generally exhibit much larger variations between different functionals and different studies.

As a case study in the influence of geometry and magnetism, we deconvolute these effects in the case of adsorption on OCUPUY, since it shows the largest deviation in adsorption energies and contains significant differences in geometry and magnetic states between functionals. 
As a most constrained system, we conducted single point calculations on PBE-optimized geometry and PBE-optimized magnetism (using the PBE-opimized wavefunction as an initial guess) with different functionals (PBE-optimized refers to PBE-D3-optimized in this section). 
In addition, to solely evaluate the effect of different magnetization in adsorption energy, single point calculations on PBE-optimized geometries without using the PBE-optimized magnetism, but setting $+1 \mu_B$ to all atoms (the default setting of VASP when running a spin-polarized calculation) are conducted.
The results are shown in \Cref{fig:adsorption_OCUPUY_fix} along with the original adsorption energy result from \Cref{fig:adsorption}. 
In addition, the difference in adsorption energy, relative to the most constrained ``PBE Geometry \& PBE Magnetism'' system is provided in \Cref{tab:OCUPUY_geomag}. 
When geometry and magnetic moment are fixed, the variation between non-hybrid functionals is similar to what is seen in metals and metal oxides, with a standard deviation of \SI{0.326}{eV}. 
The effect of magnetism alone is relatively minor in OCUPUY, with the standard deviation between functionals increasing to  \SI{0.350}{eV} when magnetic moments are relaxed, although in certain cases such as \ce{N2H} the influence can be much larger for some functionals. 
Relaxing the geometric constraints is the dominant effect in the variation between functionals, leading to a total standard deviation of \SI{0.662}{eV} between non-hybrid functionals. 
In the case of \ce{N2H} adsorption in BEEF-vdW, the difference between original adsorption energy and the one from fixed PBE geometry and PBE magnetism reaches nearly $\sim$\SI{1.5}{eV}. 
SCAN-D3 shows an even larger discrepancy when relaxing geometry, which arises due to the different unit cell geometry caused by the different orientation of the water molecule. 
In summary, systems like MOFs with many geometric degrees of freedom and multiple possible magnetic states are likely to exhibit very large differences in adsorption energies computed with different functionals, due largely to the indirect effects of different magnetic and geometric ground-states. 

\subsection{\label{sec:nitrogen}Assessment of Functionals for Nitrogen Catalysis}

While many results from this work hint at general trends in accuracy of exchange-correlation functionals, the primary scope is to evaluate functional choice for nitrogen reactions. 
Unfortunately, there is no clear  ``best choice'' of functional, and the selection will need to be based on a tradeoff of cost, accuracy, specificity, and generality. 
If a general-purpose functional is needed that works across many nitrogen reactions and catalyst types, then B3LYP-D3 is a strong choice if hybrid functionals are not too expensive. 
It demonstrates low error across all gas-phase reactions (unlike HSE06-D3 and PBE0-D3 which have errors $\sim$0.4 eV for \ce{NH3} reactions), good solid-state lattice constants, and competitive adsorption energies compared to RPA. 
If studying ammonia or \ce{NO} synthesis from \ce{N2} on metals, the BEEF-vdW functional is an excellent choice that gives strong gas-phase reaction energies and adsorption energies, but if other reactions such as \ce{NH3} oxidation are of interest, then gas-phase errors become problematic. 
The SCAN-D3 and rev-vdW-DF2 functionals are also good general-purpose choices with comparable accuracy across most metrics. 
Despite its common use, the PBE-D3 functional exhibits relatively high errors for properties other than adsorption on oxide materials, and the RPBE-D3 functional shows the largest overall errors, although this is attributed to D3 parameters that are not tuned for adsorption.

There are a few specific nitrogen adsorption systems that also warrant discussion. 
Amine-functionalization of MIL-125 is common in \ce{CO2} capture \cite{kim2013adsorption, friebe2016nh2, anjum2016mil} and has been reported to be an active photocatalyst for nitrogen fixation \cite{huang2020toward}.
Our adsorption energy results show that B3LYP-D3 has the smallest error, while the other two hybrid functionals (PBE0-D3 and HSE06-D3) and vdW functionals (rev-vdW-DF2 and BEEF-vdW) also exhibit small error.
Regarding the accuracy of gas adsorption energy in MIL-125, B3LYP-D3 appear to be the best choices among hybrid functionals because accuracy of HSE06-D3 and PBE0-D3 in the ammonia synthesis reaction is low, so B3LYP-D3 may be a better choice since it exhibits reasonable accuracy in both adsorption energy and reaction energies. 
In addition, rev-vdW-DF2 and BEEF-vdW can be accurate because of the small error in adsorption energy (RMSE of \SI{0.261}{eV} and \SI{0.238}{eV} respectively) and ammonia synthesis reaction (error of \SI{-0.055}{eV} and \SI{0.109}{eV} respectively).
Another important case is \ce{TiO2}, which has been extensively studied for photocatalytic ammonia synthesis \cite{Comer2018, comer2021computational, tian2023screening, xie2019probing, zhao2019tuning, li2022n}.
In most cases, the BEEF-vdW adsorption energies are similarly accurate to hybrid functionals, although there is a substantial deviation for \ce{CN} adsorption, where the error of BEEF-vdW and other GGA functionals is $\sim$\SI{1}{eV}. 
This suggests that hybrid calculations may be necessary for evaluating carbon-assisted pathways \cite{comer2018role, huang2023formation}, and that the HSE06-D3 or PBE0-D3 adsorption energies of NH$_{\mathrm{x}}$ species can be considered accurate despite the large error in the gas phase ammonia synthesis reaction.

There are also several notable observations about \ce{NO}. 
As discussed in \ref{sec:molecularreac}, the reaction energy for \ce{NO} synthesis from \ce{N2} and \ce{O2} is remarkably insensitive to functional choice, and is at or close to chemical accuracy for any selected functional.
This is surprising, given the complexity of the electronic structure of the \ce{O2} and \ce{NO} molecule. 
However, adsorption energies of \ce{NO} exhibit much larger errors and variations, so functional choice is still important for the \ce{NO} synthesis reaction. 
\ce{NO} adsorption energy errors on \ce{Pd} and \ce{Cu} are $\sim$ 1 eV on average, with a significant spread (std dev = \SI{0.366}{eV}) between functionals. 
The situation is somewhat better for other materials, but large deviations between functionals are observed. 
This, along with the well-known challenges in treating \ce{O2} \cite{norskov2004origin}, indicates that the accuracy of gas-phase NO synthesis is not due to an accurate treatment of \ce{NO} or \ce{O2} by the functionals, but rather a surprisingly consistent cancellation of error in all cases. 
The electronic structure origins of this phenomenon are beyond the scope of this study, but may yield useful insight into nitrogen oxidation or exchange correlation design in future work.

\subsection{\label{sec:limitations}Limitations of Dataset and Methods}

This study provides deep, but narrow, insight into the choice of exchange correlation functionals for nitrogen chemistry in heterogeneous catalysis. 
The dataset is one of the most diverse available for comparison of adsorption energies at the RPA level of theory, but it is very sparse compared to the possibilities of adsorbates and catalyst materials, and it is intentionally biased toward nitrogen chemistry. 
It is difficult to draw general conclusions from the results of the six materials and seven adsorbates studied here, especially given the chemical diversity of the materials involved. 
It is possible that the materials selected are not actually representative of metals, metal oxides, or MOFs, and hence additional tests should be performed on any given system of interest before drawing strong conclusions about the accuracy of a particular exchange correlation functional. 
However, some clear trends emerged from the small and diverse dataset studied here, and even the results from the handful of selected materials provide insight into some significant pitfalls and unexpected outcomes in the performance of exchange correlation functionals in nitrogen chemistry.

The lack of a well-established ground truth for the adsorption energies is also a limitation of this work. 
Here, we selected the optRPA level of theory as a ground truth based on the level of physics included and its strong performance in gas-phase energetics. 
However, we note that, to our knowledge, the particular strategy of optimizing the correlation contribution has not been previously applied, so while the gas-phase results are promising, the approach is not well-tested outside the scope of the gas-phase reactions and atomization energies studied here. 
The effect of the optimization of the RPA correlation can be estimated by comparing the RPA@PBE0 and optRPA results, and is relatively small in most cases, so we expect that the transferability of optRPA from molecules to solids is generally reliable. 
Another possible source of bias is the selection of wavefunctions used in optRPA, which are closely related to the PBE0 wavefunctions. 
This may bias the accuracy of hybrid functionals toward PBE0, although the modification of the exact exchange contribution from 25\% to 50\% is significant, and the results of the atomization energy benchmarks suggest that the amount of exact exchange included is more significant than the functional choice (see \ref{fig:rpa_atomization}). 
Fully self-consistent RPA correlation is required to evaluate this effect, but is beyond the scope of this study.

An additional limitation is that the RPA calculations are far more involved than standard DFT simulations. 
This human cost, along with computational cost, is illustrated in \Cref{fig:SIcost}. 
RPA calculations require testing convergence of parameters like electronic smearing that are less important in standard DFT. 
The RPA correlation is calculated separately and converges differently from the exact exchange, necessitating a careful combination of calculations performed with different settings. 
Details of all RPA calculations are provided in the SI, and we expect that they are converged with a numerical error of $\sim$0.05 eV on adsorption energies. 
Notably, this required extreme values of k-point densities and electronic smearing, especially in the case of metals, leading to massive computational costs. 
Failure to properly converge any of the numerous parameters involved in RPA can easily lead to numerical errors that exceed \SI{0.5}{eV} (see \Cref{fig:rpa_conv_pd}), so it is possible that RPA numbers obtained from the literature (or even those in this work) may have uncontrolled numerical errors. 
However, strong agreement with gas-phase properties and extensive convergence testing provides confidence that the numbers included here are sufficiently accurate to be considered a ground truth.
Notably, we find that the computational cost of RPA single point calculations is not prohibitive for typical adsorption systems, but memory requirements are higher, and significantly more convergence testing and manual analysis is required to achieve reliable results, as illustrated in \Cref{fig:SIcost}.
It is also worth noting that RPA using wavefuctions from global hybrid functionals is numerically more stable compared to the RPA@PBE approach.
This indicates that optimized RPA can be selected when higher accuracy is needed than hybrid functionals, but extra caution in choosing parameters for the calculation to obtain reliable and converged results.

\section{\label{sec:conclusion}Conclusion}

In the present work, we benchmarked multiple functionals from vairous categories (GGA, meta-GGA, hybrid, RPA) for nitrogen chemistry, using gas- and solid-phase properties. 
The results indicate that gas-phase errors for nitrogen reactions can be very large ($>$0.4 eV) even for some hybrid functionals, but that with a few minor modifications it is possible to achieve near-chemical accuracy in gas phase properties at the RPA level of theory. 
We utilized this optimized version of RPA to provide highly accurate benchmark adsorption energies of nitrogen species across a diverse range of solid materials: metals, metal oxides, and MOFs. 
The results reveal that, although standard deviations of adsorption energies between functionals fall roughly in the 0.05 - 0.30 eV range, the RMSEs when compared to accurate benchmarks can be considerably larger ($>$0.6 eV). 
Surprisingly, the adsorption energy errors were lowest on average for the oxide materials, despite the presence of highly localized electrons and van der Waals layers. 
The results generally reveal tradeoffs between the functionals studied, with B3LYP-D3 demonstrating a good compromise between gas, solid, and adsorption properties across all systems studied. 
At lower levels of theory, the BEEF-vdW functional showed excellent performance for metals, but had large errors for gas-phase reactions or materials outside its training domain. 
The rev-vdW-DF2 functional also showed strong performance for gas, solid and adsorption properties, though, like most nonhybrid functionals, it exhibits strong overbinding on metals. 
Overall, the small number of materials studied here makes it difficult to draw strong conclusions but provides some insights into which functionals to more carefully evaluate for a given material and reaction in nitrogen catalysis.

The work also revealed several interesting insights that should be generally considered when evaluating adsorption energies with DFT. 
Empirical D3 corrections can vary wildly between different functionals and in several cases lead to dispersion contributions that exceed \SI{1}{eV}, even for small and strongly chemisorbed molecules. 
This is particularly true for RPBE, where the D3 correction takes it from one of the weakest binding to one of the strongest binding functionals.
Additionally, we find large deviations in the adsorption energies of MOFs when open metal sites are present. 
These deviations can be largely attributed to changes in geometry, indicating that the flexibility of MOFs makes calculated adsorption energies more sensitive to functional choice. 
The presence of different spin states in open metal sites also contributes to these deviations, but is less significant than the geometry in the case studied here. 
However, in the case of MIL-125, small deviations in adsorption energies (standard deviation of \SI{0.140}{eV}) were commonly observed between functionals, and RMSE of \SI{0.270}{eV} was observed compared to the optimized RPA results. 
Relatively little effort has been applied to using RPA or other double-hybrids on MOFs due to their large system sizes, but these results suggest that further benchmarking and analysis is needed to reliably determine when standard DFT functionals can accurately predict the adsorption of nitrogen species in MOFs with open metal sites.

The present work provides practical insight into the variations expected between different functionals when studying nitrogen catalysis, and sheds light on the effects of multiple artifacts that researchers will encounter by following the standard workflow of adsorption energy calculations in studies of heterogeneous catalysts. 
Even with the small number of materials and adsorbates studied, it is clear that all available functionals have some weaknesses, and that care should be taken in selecting a functional if quantitative accuracy is desired.
We hope that this work serves as a useful guide for computational researchers in nitrogen catalysis who are faced with the problem of functional selection.

\section*{Supporting Information}

Additional data and computational results.
VASP simulation input and output files with sample scripts to reproduce are available in Zenodo (\href{https://doi.org/10.5281/zenodo.11506492}{10.5281/zenodo.11506491}).

\section*{acknowledgement}

N.-K.Y. and A.J.M. were funded by the U.S. Department of Energy, Office of Science, Office of Basic Energy Sciences, Chemical Sciences, Geosciences, and Biosciences Division under Award Number DE-SC0016486.
N.T. and A.J.M. were supported by funding from the National Science Foundation, under award number 1943707.
H.K. thanks Prof. Jihan Kim for supporting his visiting research stay at Georgia Tech.
We thank Prof. Phanish Suryanarayana for helpful discussions about RPA.

\printbibliography

\newpage

\section*{\label{sec:SI}Supporting Information}

\renewcommand{\thefigure}{S\arabic{figure}}
\renewcommand{\thetable}{S\arabic{table}}
\renewcommand{\theequation}{S\arabic{equation}}
\renewcommand{\thesection}{S\arabic{section}}
\setcounter{figure}{0}
\setcounter{table}{0} 
\setcounter{equation}{0}
\setcounter{section}{0}

\section{\label{sec:SI_comp_details}Computational Details}

\subsection{DFT Details}

\paragraph{Initial Position of Adsorbates}
For \ce{Cu}(100) and \ce{Pd}(111), we have two experimental adsorption energies reported, \ce{Cu}(100)+\ce{NH3} and \ce{Pd}(111)+\ce{NO} \cite{wellendorff2015benchmark}. We adopted the optimized configuration of the experimental adsorption data to be an initial binding configuration for \ce{Cu}(100) and \ce{Pd}(111). For rutile \ce{TiO2}(110), the five-fold \ce{Ti} site is chosen \cite{Comer2018}. For \ce{MoO3}(100), oxygen vacancy site is chosen \cite{Li2019}. In the case of \ce{OCUPUY}, the open vanadium site is chosen. For \ce{MIL-125}, placement of small gas molecules is less straightforward. We conducted grand-canonical monte carlo simulation of a \ce{N2} molecule and adopted the result as an initial configuration for other adsorbates. However, majority of adsorbates had too weak binding from the initial guess, which requires additional manipulation of the position by intuition. After all these processes, we were able to set initial position near the oxygen atoms bound to the metal node or the connection between the metal node and the dicarboxylate linker.

\paragraph{Correction on Experimental Unitcell Volume}
As we compare unitcell volume from DFT to one from experiment, we need to subtract the contribution from thermal expansion and zero-point effects to compare DFT-calculated volume directly to the experiment. The amount of volume change from the two factors are calculated using \cref{equ:SI_vol1} and \cref{equ:SI_vol2} \cite{lejaeghere2014error}. $\alpha_{V,rt}$ is the volume thermal expansion coefficient at room temperature, $B_0$ is the bulk modulus, $B_1$ is derivative of the bulk modulus, and $\theta_D$ is the Debye temperature.
\begin{equation}
\label{equ:SI_vol1}
       \frac{\Delta V^{thermal}}{V} \approx \int_{0}^{T_{rt}} \alpha_{V,rt}\frac{T}{T_{rt}} \,dT =\frac{\alpha_{V,rt}T_{rt}}{2}
\end{equation}

\begin{equation}
\label{equ:SI_vol2}
        \Delta V^{ZPE} = \frac{(B_1-1)\zeta}{2B_0}=\frac{9}{16}(B_1-1)\frac{k_B\Theta_D}{B_0}
\end{equation}
We used corrected volume of \ce{Cu} and \ce{Pd} from the previous benchmark from Wellendorff et al \cite{wellendorff2012density}. For \ce{TiO2} and \ce{MoO3}, we used \cref{equ:SI_vol1} and \cref{equ:SI_vol2}. $\alpha_{V,rt}$, $B_0$, $B_1$ , and $\theta_D$ are from mulitple experimental papers\cite{gerward1997post, wu1982elastic, kirby1967thermal, yamashita2021thermal, liu2009high, negishi2004anisotropic}. For MOFs, we were not able to find a full set of experimentally measured properties for correcting the volume, so the correction scheme is not applied to MOFs.

\begin{table}[p]
\setlength{\tabcolsep}{10pt}
\renewcommand{\arraystretch}{1.2}
  \caption{Three sets of different kspacing and kinetic energy cutoff used for testing convergence of energy in six solid materials.}
  \label{tab:convergence_setup}
  \begin{tabular}{lccc}
  \hline
    \hline
     & A1  & A2  & A3   \\ 
    \hline
    KSPACING (\SI{}{\per\angstrom})      & 0.5 & 0.4 & 0.3  \\
    ENCUT (eV)     & 600 & 600 & 700  \\
    \hline
    \hline
  \end{tabular}
\end{table}

\begin{table}[p]
\setlength{\tabcolsep}{8pt}
\renewcommand{\arraystretch}{1.2}
  \caption{$\Gamma$-centered k-point mesh employed to evaluate energy convergence in solid materials.}
  \label{tab:convergence_setup_kpts}
  \begin{tabular}{lcccccc}
    \hline
    \hline
    Accuracy setup & \ce{Cu}  & \ce{Pd}  & \ce{TiO2} & \ce{MoO3} & \ce{OCUPUY} & \ce{MIL-125} \\ 
    \hline
    A1      & $4\times4\times4$ & $4\times4\times4$ & $3\times3\times5$ & $4\times4\times1$ & $2\times2\times2$ & $1\times1\times1$  \\
    A2      & $5\times5\times5$ & $5\times5\times5$ & $4\times4\times6$ & $5\times5\times2$ & $3\times2\times2$ & $2\times2\times2$  \\
    A3      & $6\times6\times6$ & $6\times6\times6$ & $5\times5\times8$ & $6\times6\times2$ & $3\times3\times2$ & $2\times2\times2$  \\
    \hline
    \hline
  \end{tabular}
\end{table}

\begin{table}[p]
\setlength{\tabcolsep}{5pt}
\renewcommand{\arraystretch}{1.2}
  \caption{Convergence test result of bulk materials' energy in eV/atom unit. One GGA (PBE-D3), one meta-GGA (SCAN-D3) and one hybrid functional (PBE0-D3) are chosen for the test. The most accurate setup (A3) is set to be zero, so the numbers to be the energy difference from the A3 setup. The convergence criterion is \SI{0.025}{\eV\per atom}. }
  \label{tab:convergence_test_solids}
  \begin{tabular}{l|*{3}{c}|*{3}{c}|*{3}{c}}
    \hline
    \hline
    \multirow{2}{*}{}   & \multicolumn{3}{c|}{PBE-D3} & \multicolumn{3}{c|}{SCAN-D3} & \multicolumn{3}{c}{PBE0-D3}  \\
    \cline{2-10}
                      & A1   & A2 & A3          & A1   & A2 & A3           & A1   & A2 & A3   \\ 
    \hline
    \ce{Cu} (4 atoms)             & $-0.0033$ & $0.0318$ & $0$   & $-0.004$ & $0.0305$ & $0$   & $-0.0275$ & $0.0243$ & $0$ \\
    \ce{Pd}  (4 atoms)            & $0.0308$ & $0.0038$ & $0$   & $0.0463$ & $0.0125$ & $0$   & $0.0193$ & $-0.0193$ & $0$ \\
    \ce{TiO2} (6 atoms)          & $0.0028$ & $0.0023$ & $0$   & $0.0025$ & $0.002$ & $0$   & $-0.0053$ & $0.0002$ & $0$ \\
    \ce{MoO3}  (16 atoms)         & $0.0035$ & $0.0024$ & $0$   & $0.003$ & $0.002$ & $0$   & $-0.0046$ & $0.002$ & $0$ \\
    OCUPUY  (56 atoms)            & $0.0023$ & $-0.002$ & $0$  & $-0.0072$ & $-0.0021$ & $0$ & $0.0009$ & $ $ & $0$ \\
    MIL-125  (116 atoms)           & $0.0031$ & $0.0032$ & $0$   & $0.0034$ & $0.0034$ & $0$   & $-0.0018$ & $0.0037$ & $0$ \\
    \hline
    \hline
  \end{tabular}
\end{table}

\begin{table}[p]
\setlength{\tabcolsep}{2pt}
\renewcommand{\arraystretch}{1.2}
  \caption{Convergence test result of gas molecules' energy in eV unit. One GGA (PBE-D3), one meta-GGA (SCAN-D3) and one hybrid functional (PBE0-D3) are chosen for the test. Only $\Gamma$ point is sammpled, and the kinetic energy cutoff of \SI{400}{\eV}, \SI{500}{\eV}, \SI{600}{\eV}, \SI{700}{\eV} are tested. The most accurate setup (\SI{700}{\eV}) is set to be zero, so the numbers to be the energy difference. The convergence criterion is \SI{0.025}{\eV}, but to be aligned with the convergence test of bulk materials, \SI{600}{\eV} is chosen for the calculations of gas formation energy and gas reaction energy.}
  \label{tab:convergence_test_gases}
  \begin{tabular}{l|*{4}{c}|*{4}{c}|*{4}{c}}
    \hline
    \hline
    \multirow{2}{*}{}   & \multicolumn{4}{c|}{PBE-D3} & \multicolumn{4}{c|}{SCAN-D3} & \multicolumn{4}{c}{PBE0-D3}  \\
    \cline{2-13}
                      & \SI{400}{\eV}   & \SI{500}{\eV} & \SI{600}{\eV}& \SI{700}{\eV}         & \SI{400}{\eV}   & \SI{500}{\eV} & \SI{600}{\eV}& \SI{700}{\eV}          & \SI{400}{\eV}   & \SI{500}{\eV} & \SI{600}{\eV}& \SI{700}{\eV}     \\ 
    \hline
    \ce{N2}            & $0.051$ & $0.015$ & $0.006$& $0$   & $0.112$ & $0.015$ & $0.002$ & $0$   & $0.071$ & $0.014$ & $0.004$& $0$ \\
    \ce{O2}            & $0.030$ & $0.012$ & $0.006$& $0$   & $0.026$ & $0.013$ & $0.007$ & $0$   & $0.018$ & $0.013$ & $0.007$& $0$ \\
    \hline
    \hline
  \end{tabular}
\end{table}

\begin{table}[p]
\setlength{\tabcolsep}{5pt}
\renewcommand{\arraystretch}{1.2}
\caption{Parameters used for D3 correction.}
    \begin{tabular}{c|cccccc}
    \hline
    \hline
         &  RPBE&  PBE&  SCAN&  PBE0&  HSE06& B3LYP\\
         \hline
         $s_8$&  0.8318&  0.7875&  0.0000&  1.2177&  2.3100& 1.9889\\
         $a_1$&  0.1820&  0.4289&  0.5380&  0.4145&  0.3830& 0.3981\\
         $a_2$&  4.0094&  4.4407&  5.4200&  4.8593&  5.6850& 4.4211\\
    \hline
    \hline
    \end{tabular}
    \label{tab: d3_param}
\end{table}

\subsection{RPA Details}

\paragraph{RPA Calculation}
In RPA@PBE approach for metallic systems, the finite-temperature RPA formalism \cite{kaltak2014cubic} was utilized with Fermi smearing and a smearing width ($\sigma$) of \SI{0.05} {\eV}. For non-metallic systems, the non-finite-temperature formalism and low $\sigma$ values ($\leq$ \SI{0.005}{\eV}) were used. In RPA@PBEx calculations, which utilize global hybrid functionals such as PBE0 and PBEx50, the non-finite-temperature formalism and Gaussian smearing were applied with smearing widths of \SI{0.05}{\eV} for metals and \SI{0.005}-\SI{0.01}{\eV} for non-metals. Generally, RPA@PBE and RPA@HSE06 calculations are more unstable and sensitive to the calculation parameters, such as smearing widths and k-point meshes, compared to RPA@PBEx approaches.

\begin{figure}[p]
    \centering
    \includegraphics[width=\textwidth]{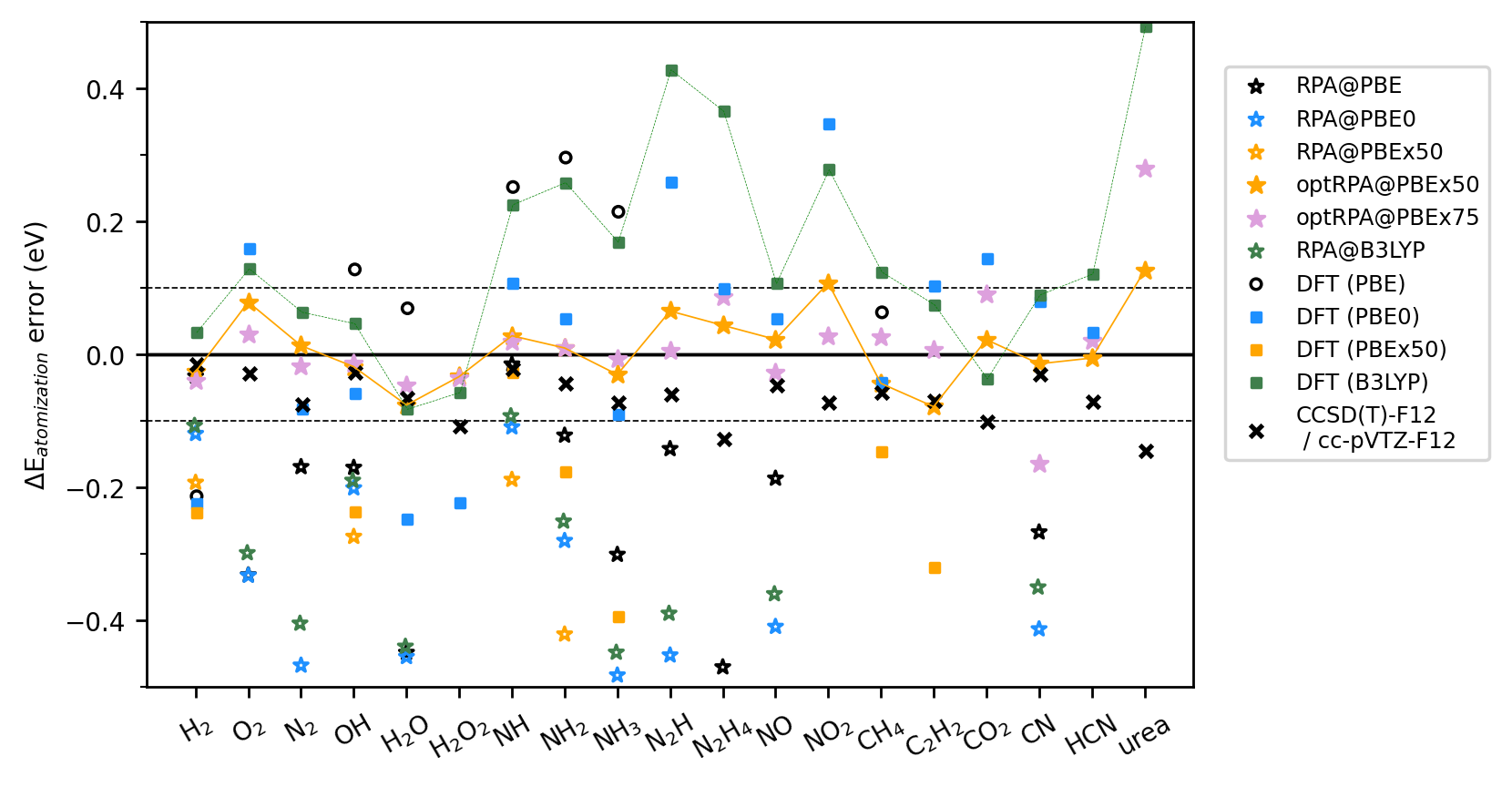}
    \caption{Error of the atomization energies from CCSD(T) values.}
    \label{fig:rpa_atomization}
\end{figure}

\begin{figure}[p]
    \centering
    \includegraphics[width=\textwidth]{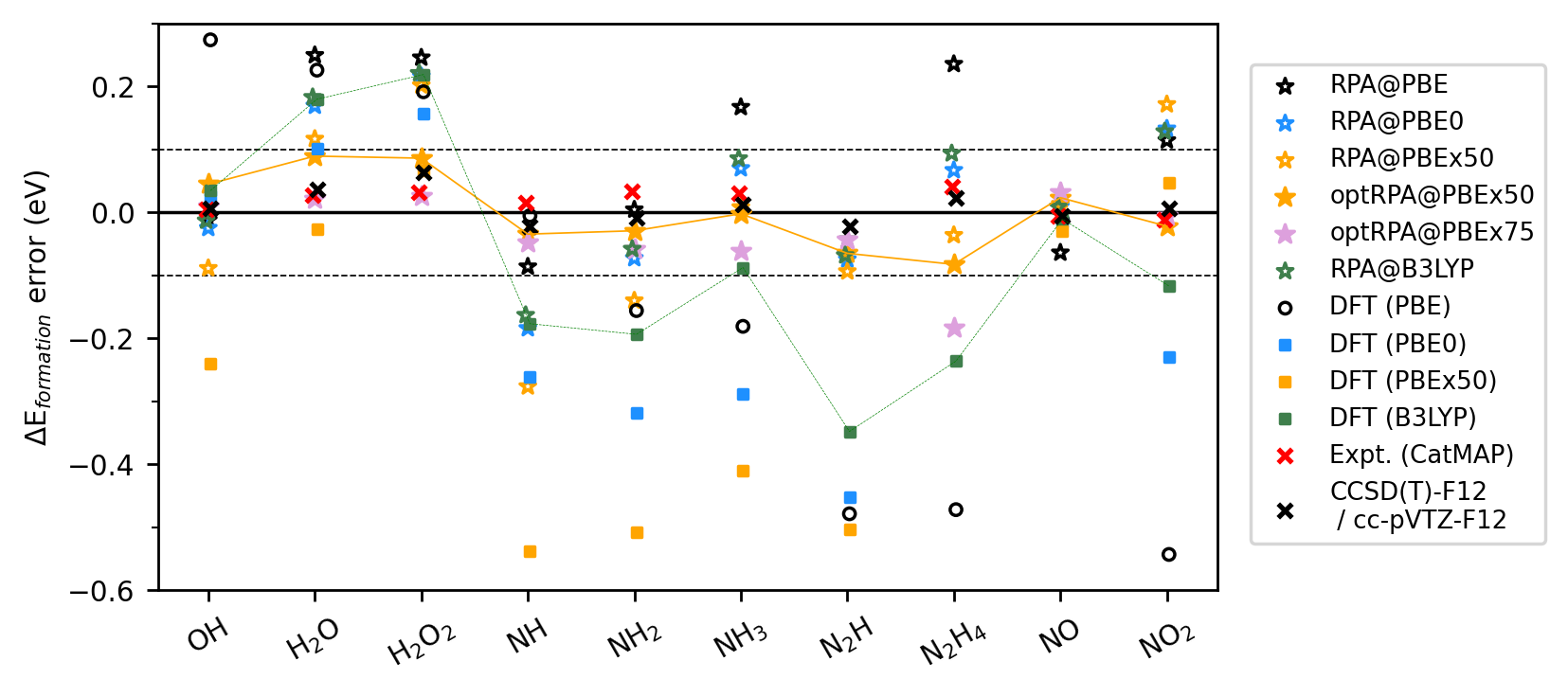}
    \caption{Error of the formation energies from CCSD(T) values.}
    \label{fig:rpa_formation}
\end{figure}

\paragraph{RPA molecular Energy}
Observations indicate that the absolute contributions of $E_{c,RPA}$ to both molecular atomization energy and formation energy are underestimated in RPA calculations. 
To address this issue, a scaling factor was multiplied to E$_{c,RPA}$. 
The scaling factor was optimized to minimize the error in atomization energy of molecules considered in \Cref{fig:rpa_atomization}, compared to CCSD(T) values. 
CCSD(T) energies were calculated using ORCA v5.0.4\cite{neese2012orca}, and the extrapolation of basis sets was applied using aug-cc-pVQZ and aug-cc-pV5Z. Unlike formation energy, atomization energy is less affected by error cancellation between the energies of reactants and products, making it a more suitable dataset for fitting purposes. Utilizing the PBE-optimized molecular structures, the scaling factors was determined to be 1.17 for the optRPA@PBEx50 approach. The error of optRPA@PBEx50 is comparable to the numerical error in CCSD(T), as can be seen in the error of CCSD(T)-F12, a CCSD(T) method with F12 correction, which improves the basis set dependency. \cite{thomas2007f12}
The same scaling approach would not work for RPA@PBE as we have seen examples where the signs of $E_{HF}$ and $E_{c,RPA}$ of molecules were not correct. 
In the assessment of the sensitivity of the $E_{c,RPA}$ scaling factor using k-fold cross-validation, we observed consistent values for the scaling factor in different fold configurations. 
For the optRPA@PBEx50 approach, considering k=3, 5, and 10 folds, the mean (median, standard deviation) of the optimized scaling factors were found to be 1.168 (1.168, 0.0002), 1.167 (1.168, 0.0010), and 1.167 (1.168, 0.0007), respectively. 
The scaling factor optimized based on the formation energy was identified as 1.174. When employing B3LYP-optimized structures and atomization energies, and the scaling factors were determined to be 1.176 for optRPA@PBEx50 and 1.248 for optRPA@PBEx75. B3LYP-optimized structures were used to obtain the energies for \Cref{fig:rpa_atomization} and \Cref{fig:rpa_formation}, but deviations when PBE geometries are used were observed to be minimal in all cases tested.

\begin{table}[p]
\setlength{\tabcolsep}{12pt}
\renewcommand{\arraystretch}{1.2}
  \caption{Convergence test results of RPA molecular atomization energy in eV unit. The most accurate setup (\SI{700}{eV} and \SI{600}{eV} for $E_{HF}$ and $E_{c,RPA}$, respectively) is set to be zero, so the numbers to be the energy difference.}
  \label{tab:rpa_conv_mol}
  \begin{tabular}{l|*{4}{c}}
    \hline
    \hline
    \multirow{2}{*}{}   & \multicolumn{4}{c}{RPA@PBE0}  \\
    \cline{2-5}
                      & 600/\SI{400}{\eV}  & 600/\SI{500}{\eV}
                      & 600/\SI{600}{\eV} & 700/\SI{600}{\eV}  \\ 
    \hline
    \ce{N2}            & $0.098$ & $0.023$ & $0.002$ & $0$   \\
    \ce{O2}            & $0.028$ & $0.024$ & $0.028$ & $0$   \\
    \hline
    \hline
  \end{tabular}
\end{table}

\paragraph{RPA Energy Convergence}

\begin{figure}[p]
    \centering
    \includegraphics[width=\textwidth]{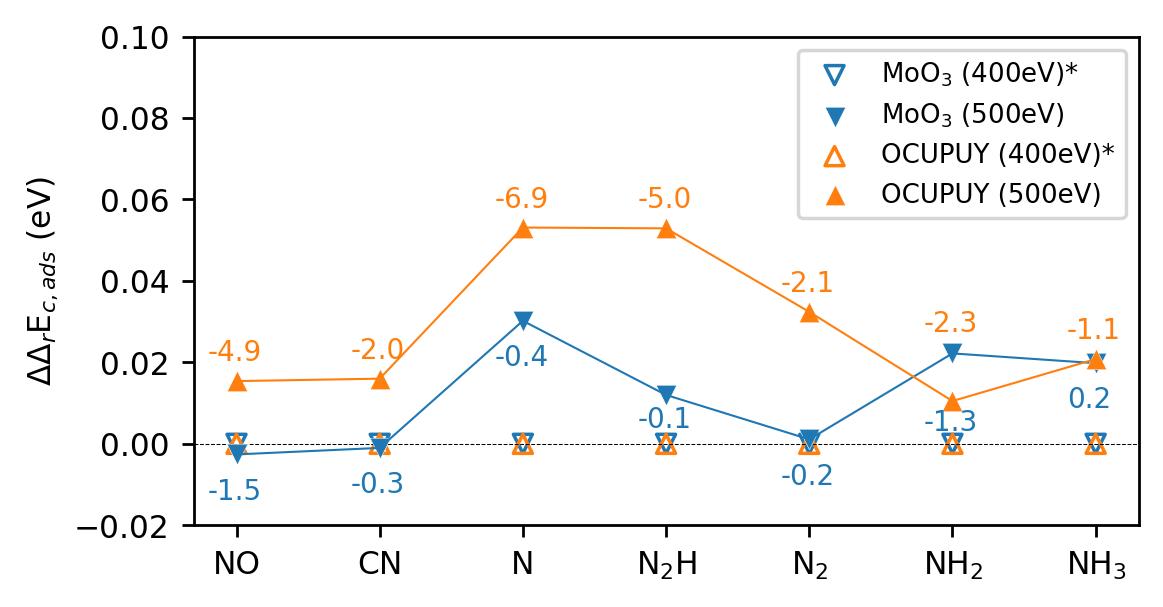}
    \caption{Convergence test results of E$_{c,RPA}$ contribution to $\Delta_r$E$_{ads}$ in terms of ECUT. *denotes the parameters used in the main text.}
    \label{fig:rpa_conv_ecut}
\end{figure}

Detailed convergence test results are provided for RPA calculations across various systems. RPA calculations are computationally intensive, especially in terms of memory requirements. Therefore, a lower kinetic energy cutoff (ENCUT) of \SI{400}{\eV} was chosen for calculating $\Delta_r$E$_{c,ads}$ as defined below, which represents the E$_{c,RPA}$ contribution to the adsorption energy $\Delta_r$E$_{ads}$, after a convergence test (\Cref{fig:rpa_conv_ecut}). 
\begin{equation}
    \label{equ:ads_r}
    \Delta_rE_{ads} = E_{solid+gas}-E_{solid}-E_{gas}
\end{equation}
Among the GW pseudopotentials used, the oxygen (O\_GW\_new) and nitrogen (N\_GW\_new) pseudopotentials have the highest default ENCUTs of 434.4 and \SI{420.9}{\eV}. Consequently, the observation that, for \ce{MoO3} systems containing N-containing adsorbates, the differences in $\Delta_r$E$_{ads}$ between ENCUTs of 400 and \SI{500}{\eV} are less than \SI{0.02}{eV} suggests that \SI{400}{\eV} is a reasonable choice (\Cref{fig:rpa_conv_ecut}(a)). Although differences of about \SI{0.05}{\eV} are observed for OCUPUY, the absolute $\Delta_r$E$_{c,ads}$ exceeds \SI{5}{\eV}, corresponding to roughly a 1\% error (\Cref{fig:rpa_conv_ecut}(b)).

\begin{figure}[p]
    \centering
    \includegraphics[width=\textwidth]{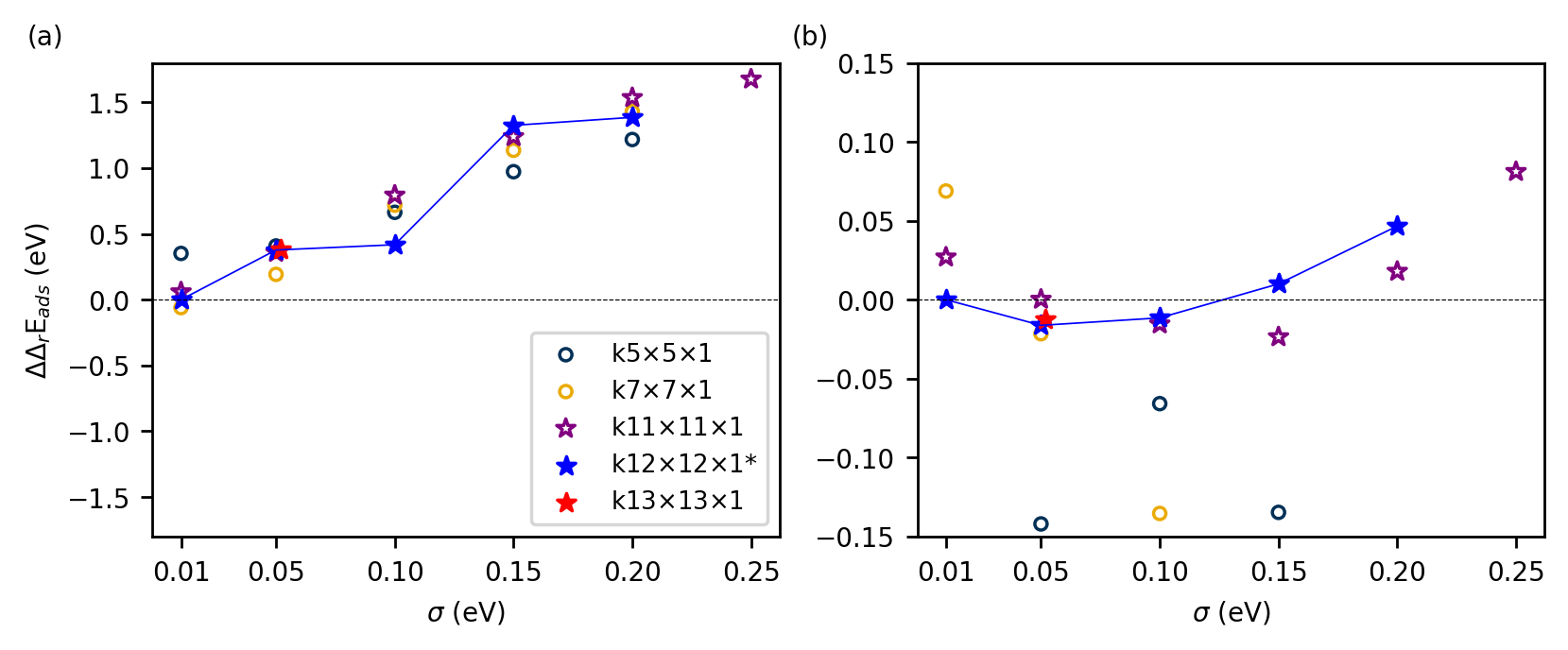}
    \caption{Convergence test of RPA@PBE calculations on Pd. (a) with $E_{HF}$ correction. (b) without $E_{HF}$ correction and after subtracting $\frac{1}{2}$T$\Delta$S$_{elec}$. * denotes the parameters used in the main text.}
    \label{fig:rpa_conv_pd}
\end{figure}

\begin{figure}[p]
    \centering
    \includegraphics[width=\textwidth]{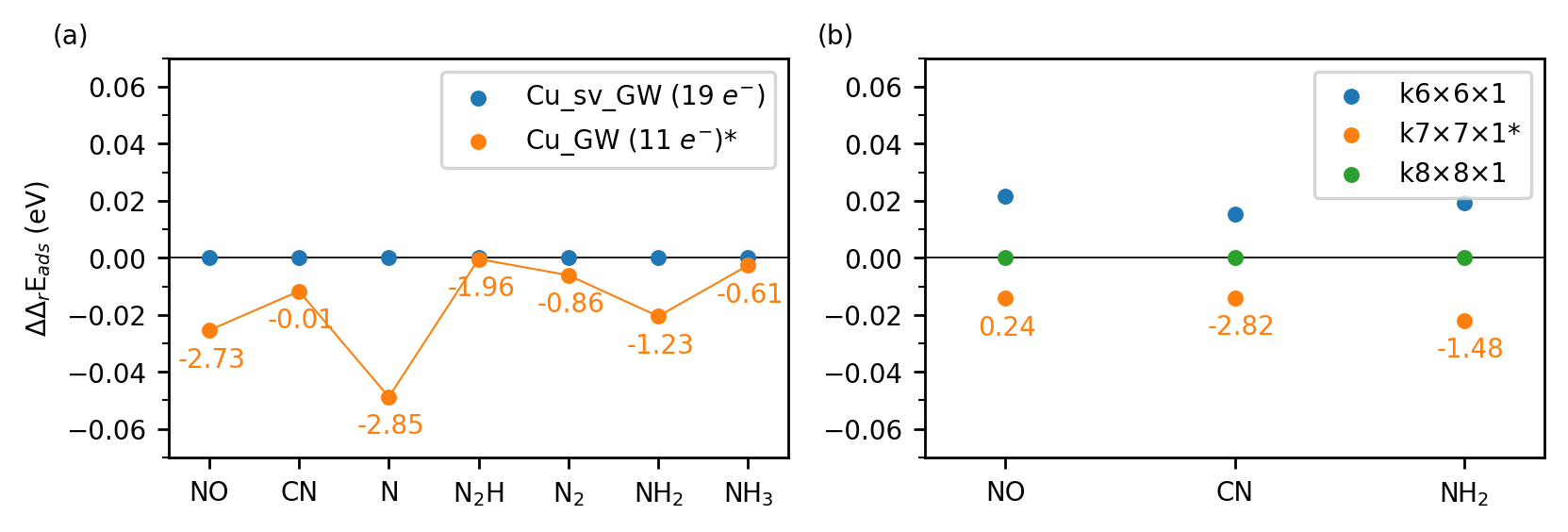}
    \caption{Convergence test results of RPA@PBE0 calculations on Cu in terms of pseudopotentials (a) and k-points grid (b). * denotes the parameter used in the main text.}
    \label{fig:rpa_conv_cu}
\end{figure}

In the RPA@PBE approach to metallic systems, the non-finite-temperature formalism may break down; thus, the finite-temperature formalism should be used. Additionally, the calculations on metallic systems are sensitive to $\sigma$, as the metallic slabs, with and without adsorbates, respond differently to variations in $\sigma$ (\Cref{fig:rpa_conv_pd}). The different responses can be somewhat alleviated by subtracting half of the electronic entropy term from $E_{HF}$, similar to how the zero-temperature energy is obtained in DFT calculation. Furthermore, a correction to $E_{HF}$ related to partial occupancies \cite{harl2010Ecorr} was not applied, as it worsens the energy convergence in terms of $\sigma$ (\Cref{fig:rpa_conv_pd}(a)). According to the VASP forum, the correction cannot be rigorously derived within the finite-temperature formalism. Therefore, the correction to $E_{HF}$ was applied only to calculations using the non-finite-temperature formalism. RPA calculations on metallic systems generally require much denser k-point grids than standard DFT.
For the metallic slabs of \ce{Cu} and \ce{Pd}, k-point grids of 7$\times$7$\times$7 and 12$\times$12$\times$12 were used, respectively (See \Cref{fig:rpa_conv_pd}(b) and \Cref{fig:rpa_conv_cu}(b) for energy convergence test). Another parameter worth discussing is the choice of GW pseudopotentials. Specifically, for Cu, a GW pseudopotential with fewer valence electrons was utilized after convergence test (\Cref{fig:rpa_conv_cu}(a)). The reason behind the choice is the high computational cost associated with generating PBEx wavefunctions, rather than the actual RPA calculation step to get E$_{c,RPA}$.

\begin{figure}[p]
    \centering
    \includegraphics[width=\textwidth]{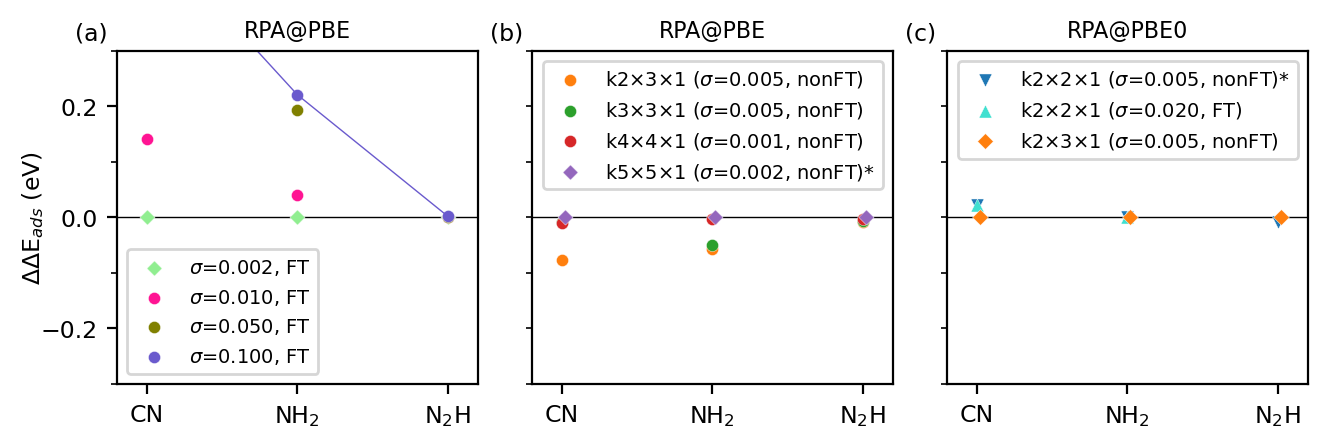}
    \caption{Convergence test results of RPA calculations on TiO$_2$. (a) RPA@PBE with finite-temperature (FT) formalism. (b) RPA@PBE without finite-temperature formalism. (c) RPA@PBE0. * deontes the parameter used in the main text.}
    \label{fig:rpa_conv_tio2}
\end{figure}

\begin{figure}
    \centering
    \includegraphics[width=\textwidth]{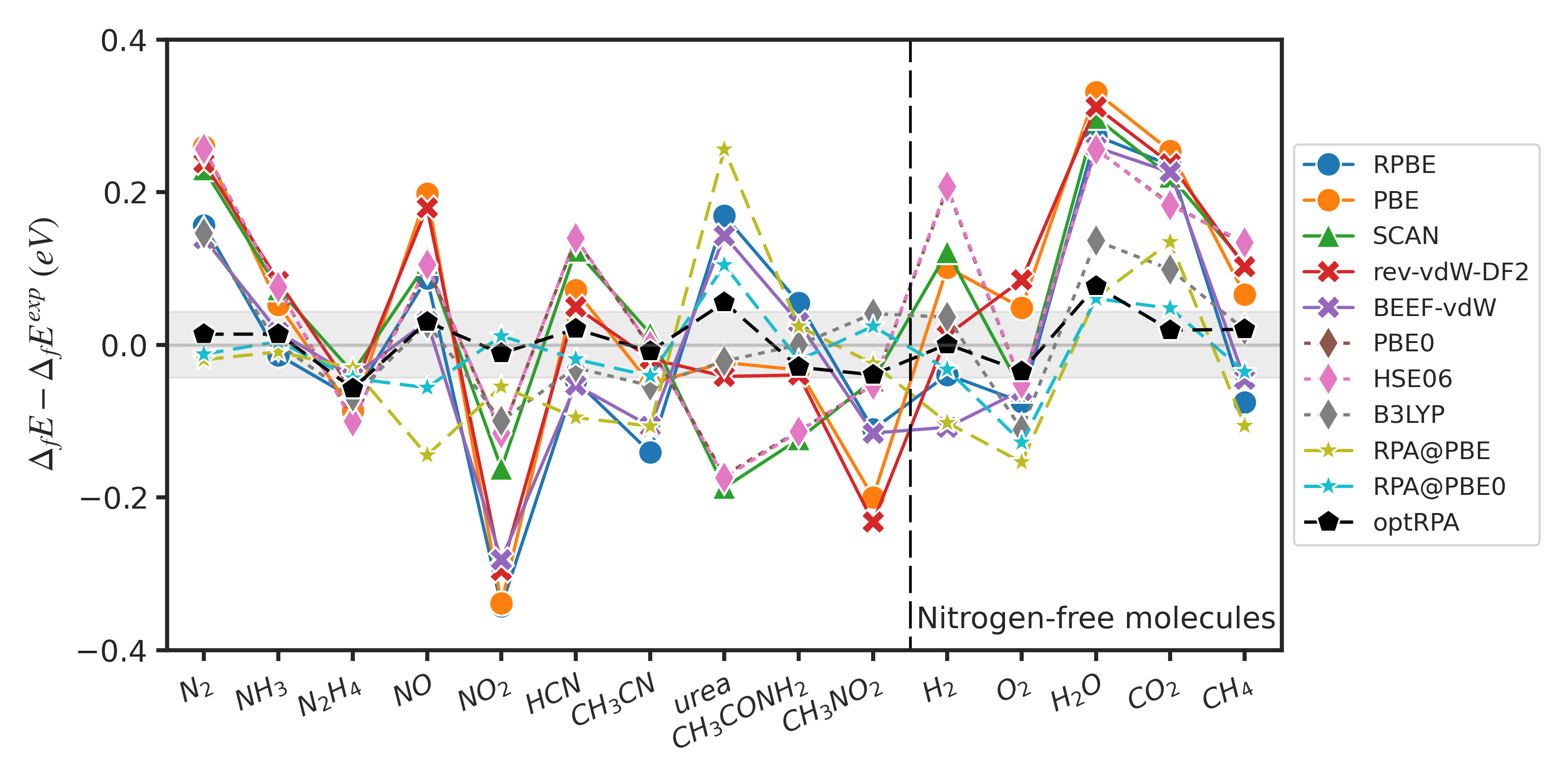}
    \caption{Error of calculated formation energy per each gas molecule without D3 correction.}
    \label{fig:SIformation}
\end{figure}

\begin{figure}
    \centering
    \includegraphics[width=\textwidth]{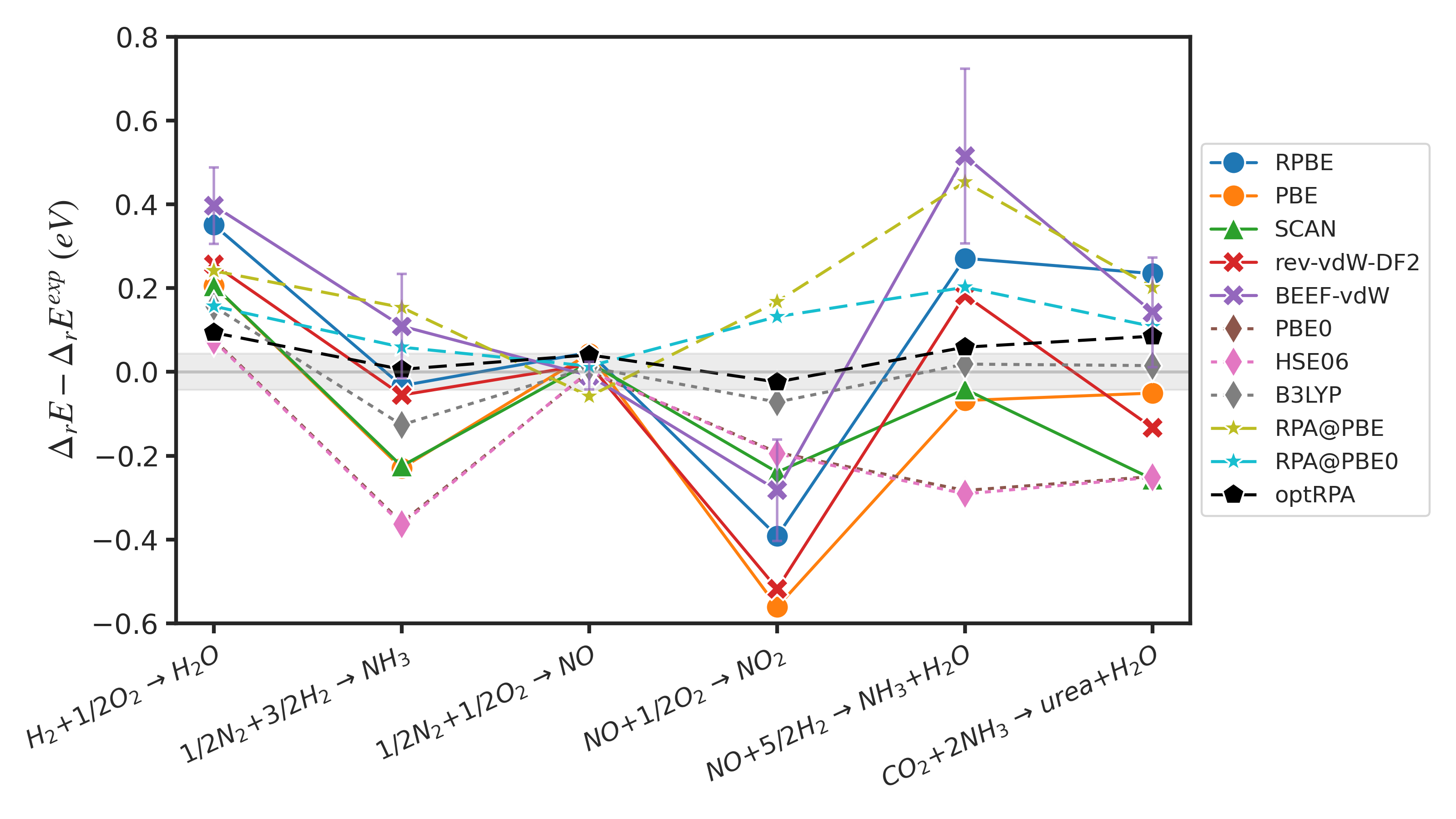}
    \caption{Error of calculated reaction energy from experimental reaction energy without D3 correction.}
    \label{fig:SIreaction}
\end{figure}

\begin{figure}
    \centering
    \begin{tabular}{cc}
        \includegraphics[width=0.5\textwidth ]{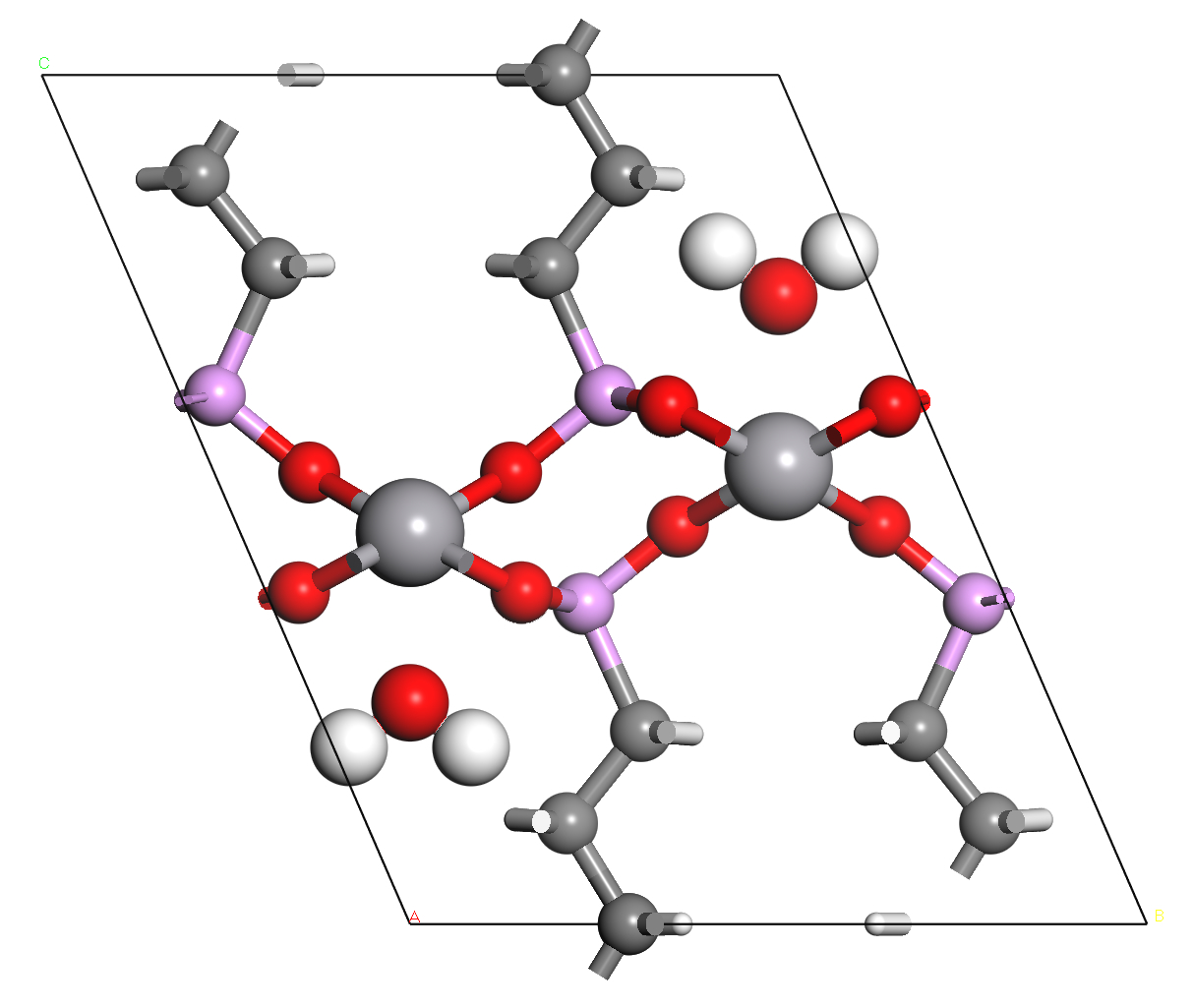}& \includegraphics[width=0.5\textwidth ]{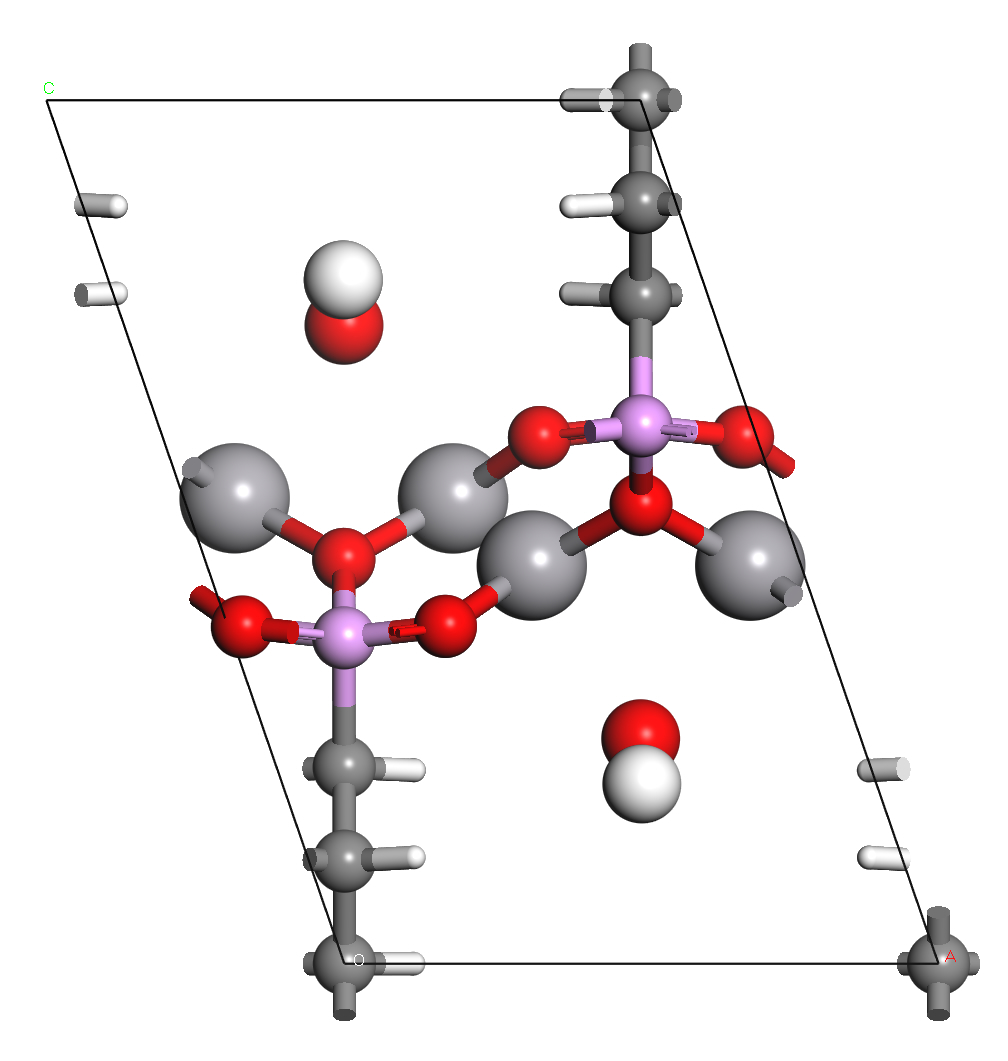}\\ \includegraphics[width=0.5\textwidth ]{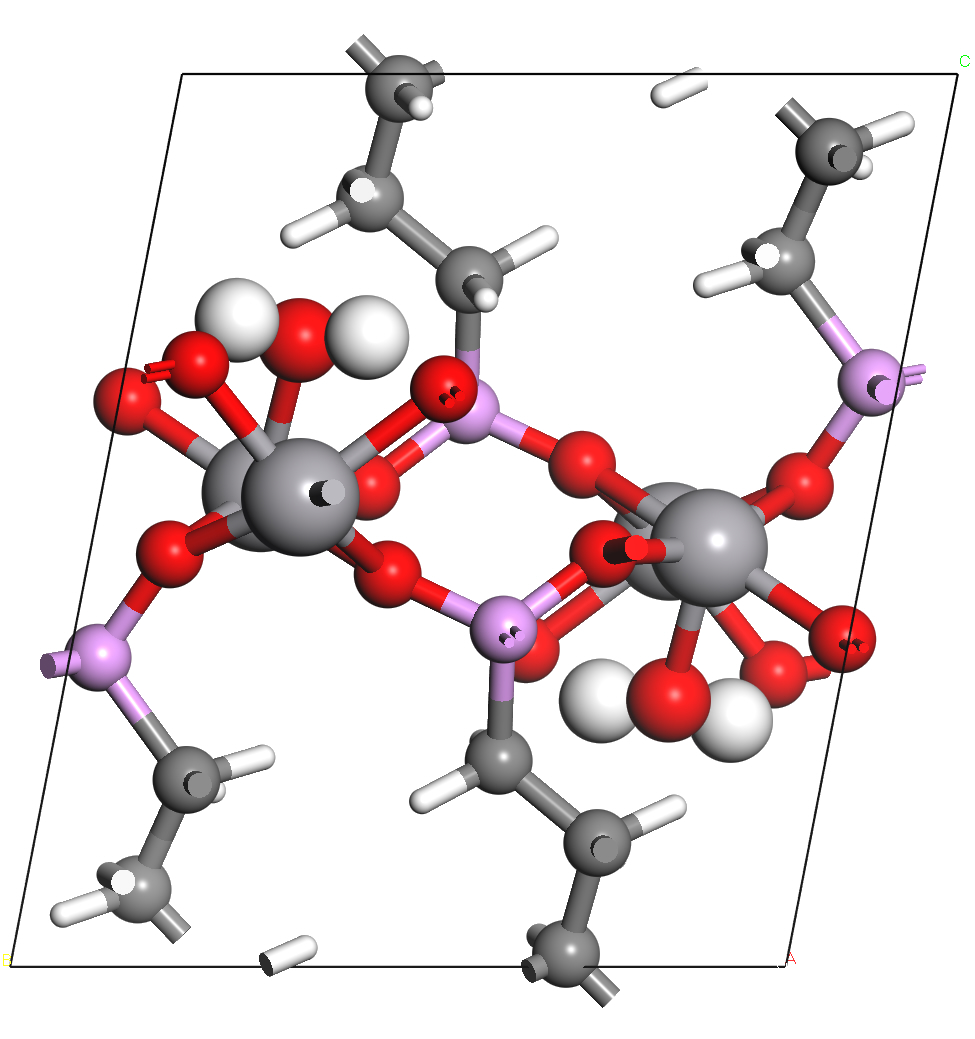}& \includegraphics[width=0.5\textwidth ]{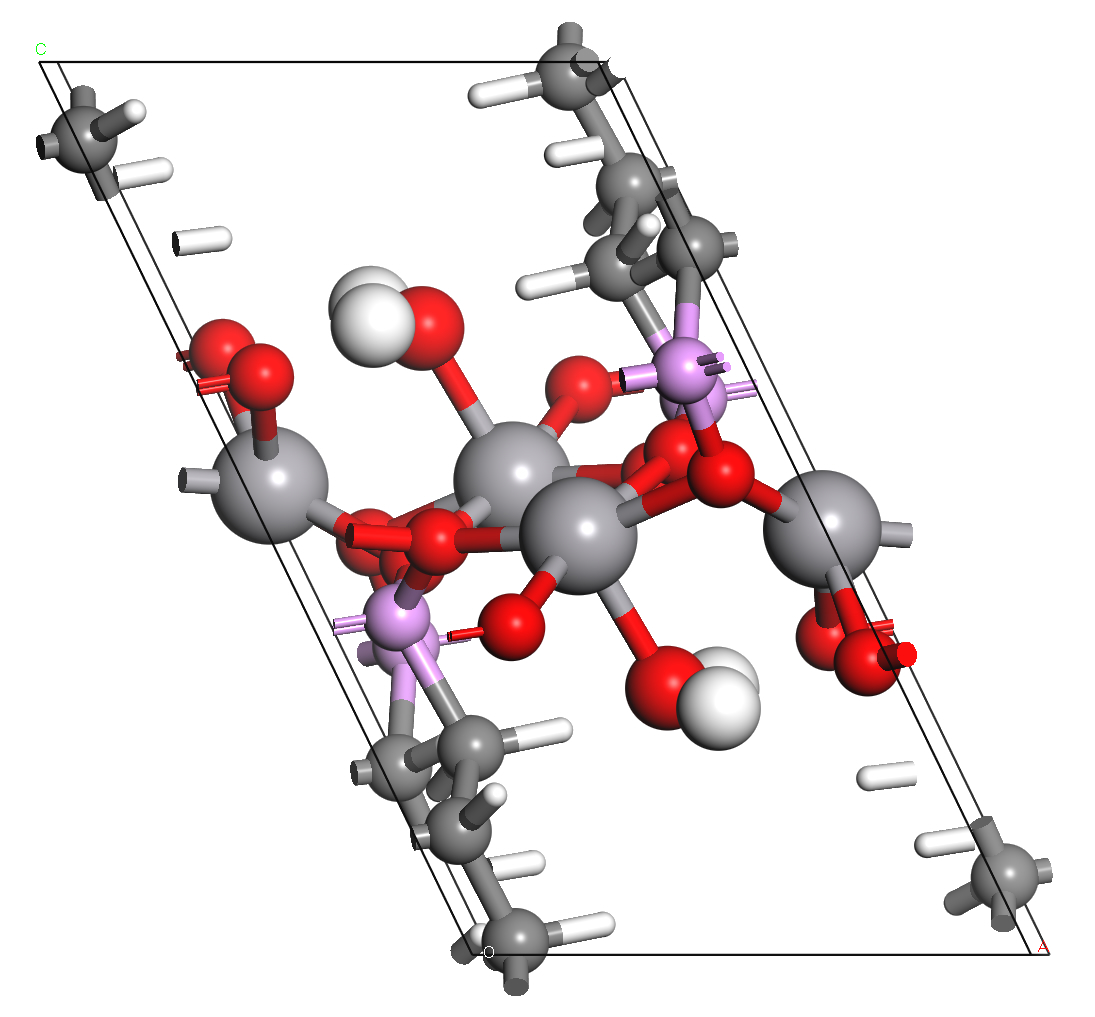}\\
    \end{tabular}
    \caption{Images of relaxed OCUPUY unitcell using (top) PBE-D3 and (bottom) SCAN-D3. (color code: light grey=V, grey=C, red=O, white=H, purple=P; vanadium atoms and water molecules are magnified in the images.)}
    \label{fig:SIOCUPUY_water}
\end{figure}

\begin{figure}
  \centering
  \begin{subfigure}
    \centering
    \includegraphics[width=\linewidth]{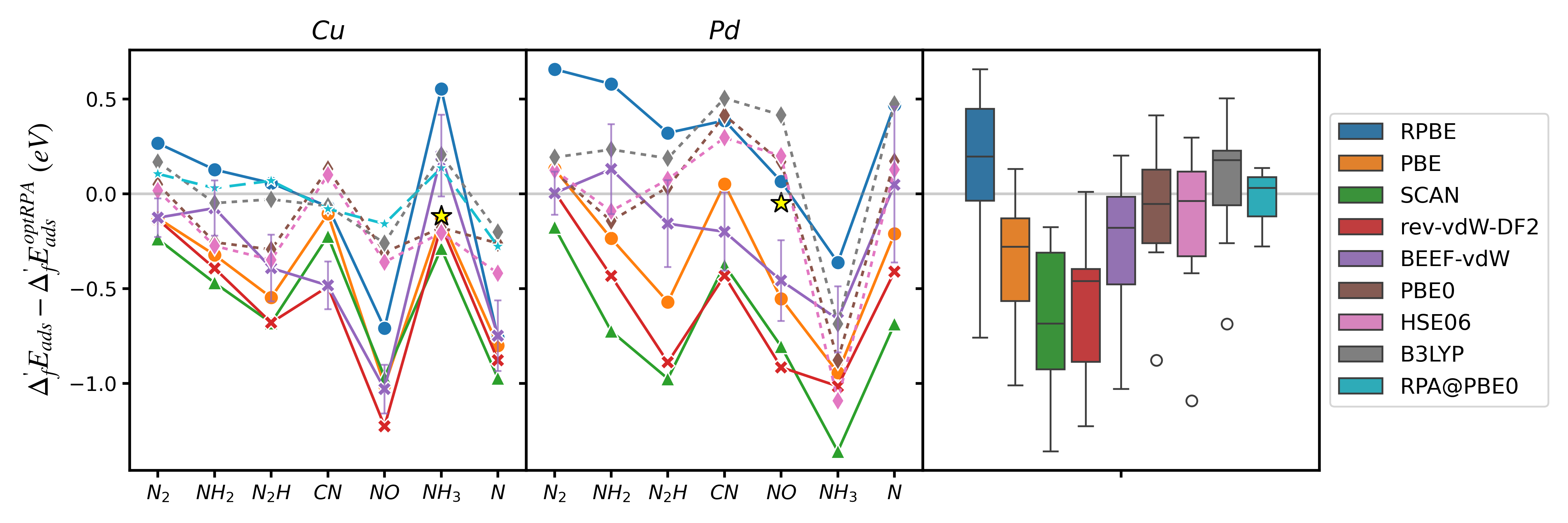}
  \end{subfigure}
  \hfill
  \begin{subfigure}
    \centering
    \includegraphics[width=\linewidth]{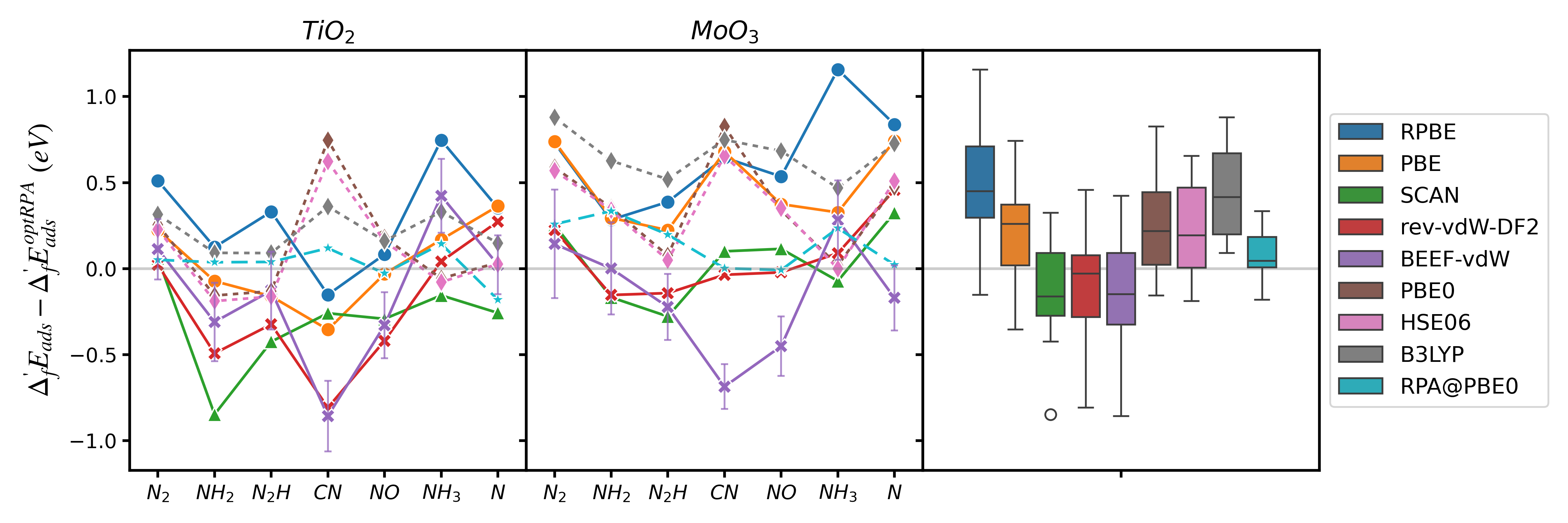}
  \end{subfigure}
  \hfill
  \begin{subfigure}
    \centering
    \includegraphics[width=\linewidth]{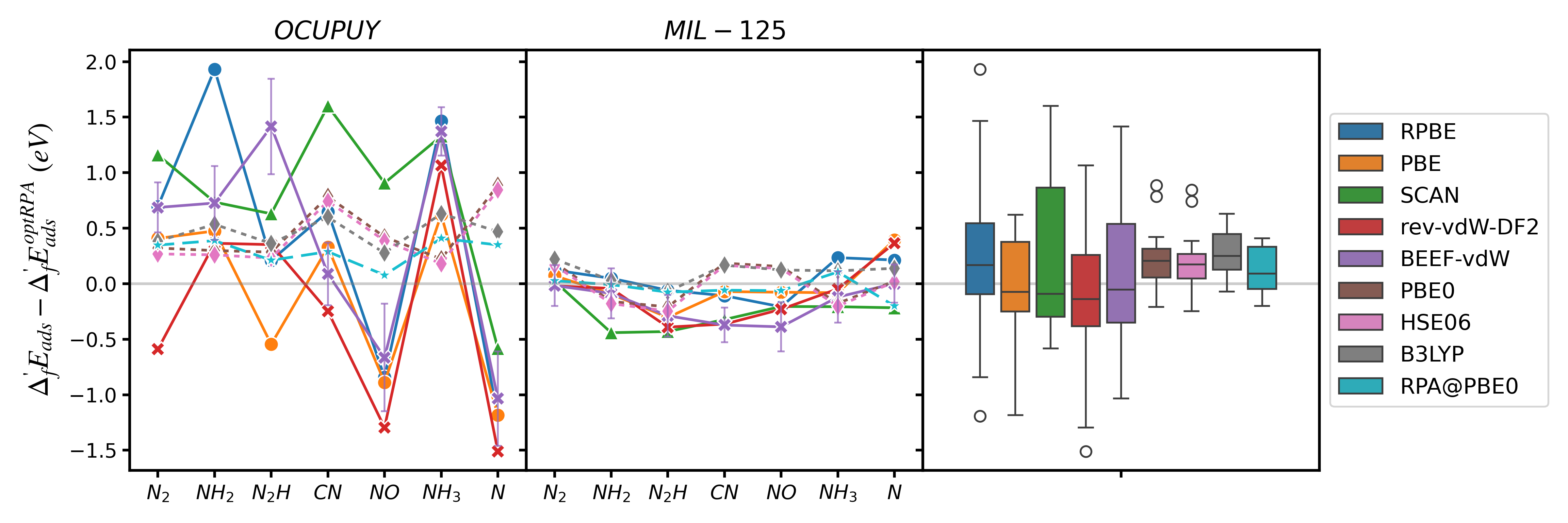}
  \end{subfigure}
  \caption{Error of the calculated adsorption energies without D3 correction in (top) metals, (center) metal oxides, and (bottom) MOFs compared to the adsorption energy of optRPA. Each point is drawn by subtracting D3 correction in Figure 4. Distribution of errors for each functional are plotted on the right as box and whisker plots.}
  \label{fig:SIadsorption_nod3}
\end{figure}

\begin{figure}
  \centering
  \begin{subfigure}
    \centering
    \includegraphics[width=\linewidth]{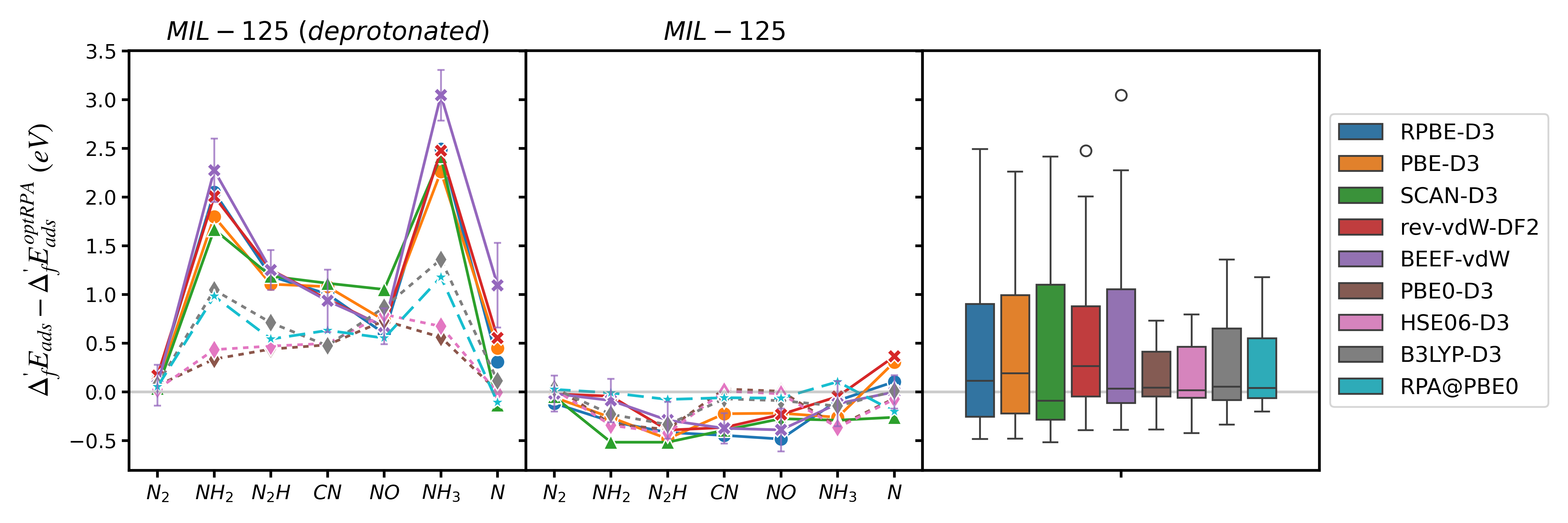}
  \end{subfigure}
  \hfill
  \begin{subfigure}
    \centering
    \includegraphics[width=\linewidth]{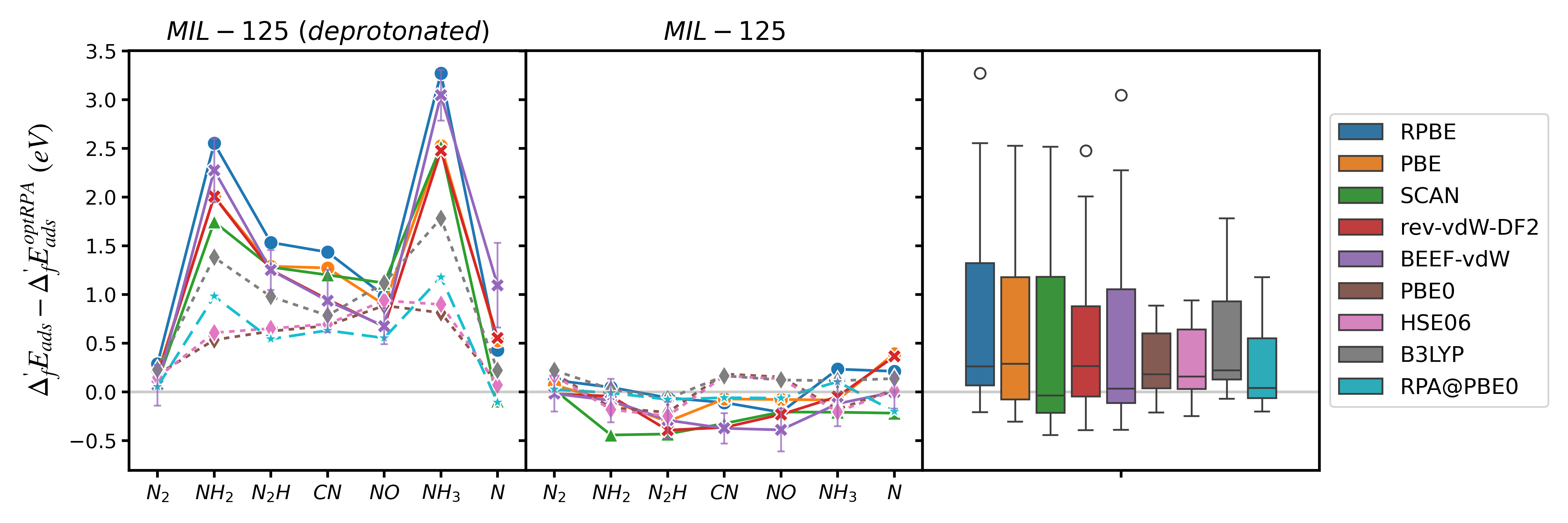}
  \end{subfigure}
  \caption{Error of the calculated adsorption energies with (top) and without (bottom) D3 correction compared to the adsorption energy of optRPA in deprotonated MIL-125 and protonated MIL-125.}
  \label{fig:SIadsorption_deprotonated_MIL125}
\end{figure}

\begin{figure}[p]
    \centering
    \includegraphics[width=\textwidth]{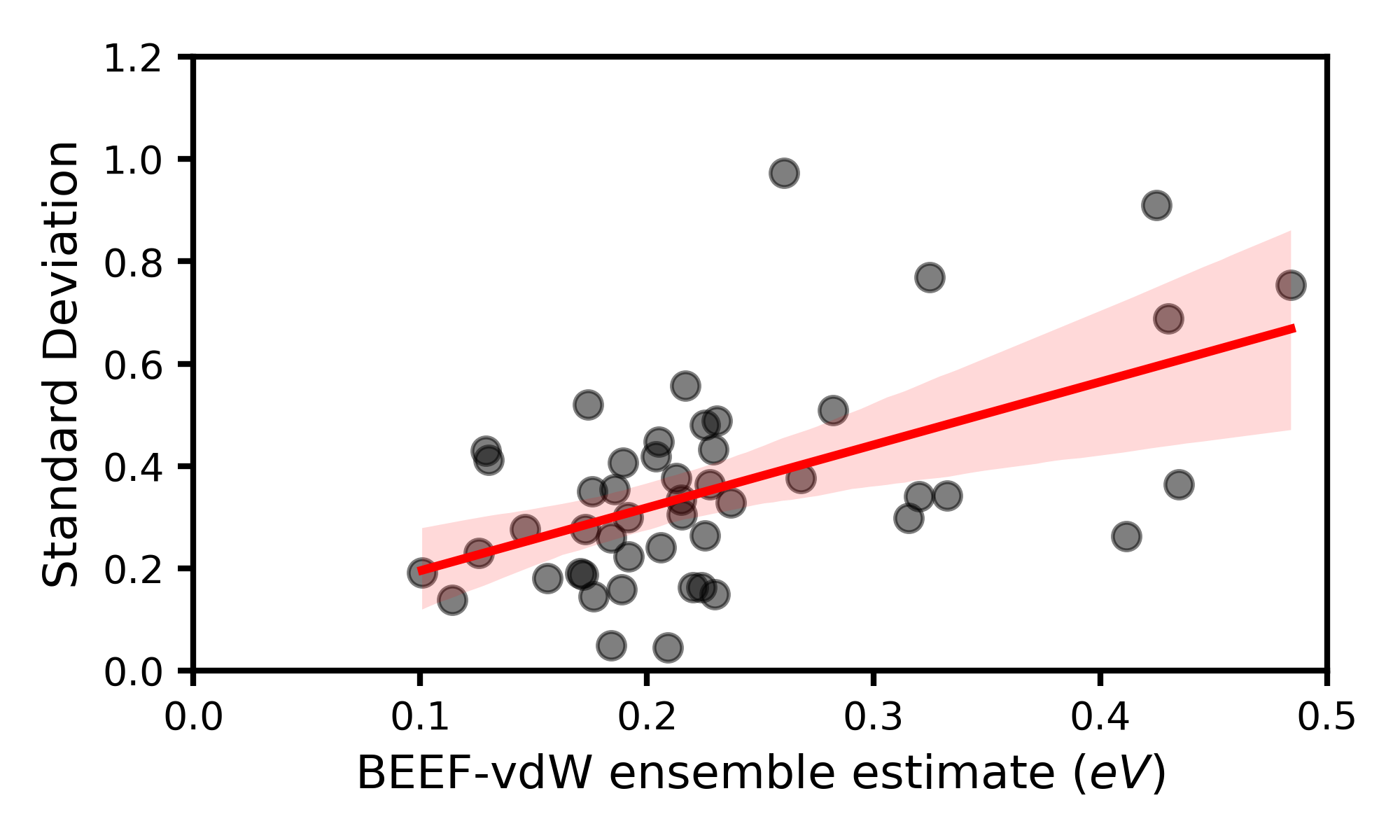}
    \caption{Relation between BEEF-vdW error bar and standard deviation of adsorption energies from different functionals. Rgression line is drawn, and confidence interval of 95\% is shaded in red. Pearson correlation coefficient is 0.536.}
    \label{fig:SIbeef}
\end{figure}

RPA@PBE calculations on \ce{TiO2} reveal a sensitivity to $\sigma$ of non-metallic systems when adsorbate-induced states are near the Fermi level, akin to observations in the metallic systems. Likewise, the sensitivity to $\sigma$ occurs due to the different responses of slabs, with and without adsorbates, to changes in $\sigma$ (\Cref{fig:rpa_conv_tio2}(a)). To minimize the smearing effect, extremely low $\sigma$ values ($\leq$ \SI{0.005}{\eV}) and the non-finite-temperature formalism were used (\Cref{fig:rpa_conv_tio2}(b)). However, it is important to note that such low $\sigma$ values can lead to numerical instability. For the RPA methods prone to instability, such as RPA@PBE and RPA@HSE06, it is important to ensure numerical stability by verifying the consistency of DFT energy before and after the exact diagonalization step. Additionally, the RPA@PBE method requires denser k-point grids compared to those required by global hybrid-based approach, likely due to the influence of states near the Fermi level (\Cref{fig:rpa_conv_tio2}(b,c)). In RPA@PBEx approaches for non-metallic systems, the same or denser k-point grids, compared to the ones used for DFT calculations, were employed, except for \ce{TiO2}, where its energy convergence is tested in \Cref{fig:rpa_conv_tio2}(c).

\newpage
\newpage

\section{\label{sec:SI_SI}Supplemenatry Figures and Tables}

\begin{table}[p]
\setlength{\tabcolsep}{4pt}
\renewcommand{\arraystretch}{1.3}
  \caption{Error of the calculated formation energies and reaction energies compared to the experimental formation and reaction energies without D3 correction.}
  \label{tab:formation_reaction_err_nod3}
  \begin{tabular}{l|*{3}{c}|*{3}{c}}
    \hline
    \hline
    \multirow{2}{*}{}   & \multicolumn{3}{c|}{$\Delta^{LS}_fE - \Delta_f E^{exp}$} & \multicolumn{3}{c}{$\Delta_r E - \Delta_r E^{exp}$}\\
    \cline{2-7}
                        & MaxError & MAE& RMSE & MaxError & MAE& RMSE   \\ 
    \hline
            RPBE        &  $-0.342$ & $0.126$ &  $0.156$ & $-0.392$ &  $0.221$&  $0.261$\\
            PBE         &  $-0.339$ & $0.141$ &  $0.177$ & $-0.561$ &  $0.193$&  $0.264$\\
            SCAN        &  $0.298$ & $0.127$ &  $0.149$ & $-0.256$ &  $0.165$&  $0.191$\\
            rev-vdW-DF2 &  $0.312$ & $0.133$ &  $0.167$ & $-0.518$ &  $0.194$&  $0.254$\\
            BEEF-vdW    &  $-0.282$ & $0.110$ &  $0.138$ & $0.515$ &  $0.242$&  $0.299$\\
            PBE0        &  $0.256$ & $0.130$ &  $0.148$ & $-0.359$ &  $0.194$&  $0.229$\\
            HSE06       &  $0.256$ & $0.131$ &  $0.149$ & $-0.363$ &  $0.196$&  $0.232$\\
            B3LYP       &  $0.146$ & $0.059$ &  $0.075$ & $0.156$ &  $0.066$&  $0.088$\\
            RPA@PBE     &  $0.256$ & $0.088$ &  $0.110$ & $0.453$ &  $0.213$&  $0.245$\\
            RPA@PBE0    &  $-0.128$ & $0.043$ &  $0.054$ & $0.202$ &  $0.112$&  $0.128$\\
            optRPA      &  $0.077$ & $0.029$ &  $0.035$ & $0.094$ &  $0.051$&  $0.060$\\
    \hline
    \hline
  \end{tabular}
\end{table}

\begin{table}[p]
\setlength{\tabcolsep}{8pt}
\renewcommand{\arraystretch}{1.2}
    \caption{Error of calculated unitcell volume per atom from the corrected experimental volume with and without D3 correction. Unit in \SI{}{\angstrom^3/atom}.}
    \begin{tabular}{l|*{2}{c}}
        \hline
        \hline
           & $V_{DFT}-V_{exp}$ with D3  & $V_{DFT}-V_{exp}$ without D3\\
        \hline
        & \multicolumn{2}{c}{PBE}\\
         \hline
         \ce{Cu}&  -0.264 & 0.430\\
         \ce{Pd}&  0.087& 0.832\\
         \ce{TiO2}&  0.248& 0.438\\
         \ce{MoO3}&  0.701& 2.215\\
         \hline
         &\multicolumn{2}{c}{RPBE}\\
         \hline
         \ce{MoO3}&  -0.181& 3.288\\
         \hline
         \hline
    \end{tabular}
    \label{tab:vol_w_wo_d3}
\end{table}

\begin{table}
\setlength{\tabcolsep}{6pt}
\renewcommand{\arraystretch}{0.93}
  \caption{Calculated adsorption energy ($\Delta_f^{'}E_{i,ads}$) from RPAs (optRPA, RPA@PBEx50, RPA@PBE0, RPA@PBE). Unit in \SI{}{eV}.}
  \label{tab:SIads_E_RPAs}
  \begin{tabular}{l|l|*{4}{c}}
    \hline
    \hline
    \multirow{2}{*}{}  & & \multicolumn{4}{c}{$\Delta_f^{'} E_{ads}$}\\
    \cline{2-6}
                        &adsorbate & optRPA & RPA@PBEx50 & RPA@PBE0 & RPA@PBE   \\ 
    \hline
        \multirow{7}{*}{\ce{Cu}}& \ce{N2} & $0.044$ & $0.186$ & $0.149$& $0.206$ \\
        & \ce{NH2} & $0.024$ & $0.112$ & $0.054$& $0.094$ \\
        & \ce{N2H} & $2.215$ & $2.369$ & $2.283$& $2.237$ \\
        & \ce{CN} & $2.749$ & $2.782$ & $2.670$& $2.374$ \\
        & \ce{NO} & $3.580$ & $3.736$ & $3.421$& $3.134$ \\
        & \ce{NH3} & $-1.321$ & $-1.163$ & $-1.184$& $-1.136$ \\
        & \ce{N} & $4.158$ & $4.106$ & $3.882$& $3.593$ \\
        \hline
        \multirow{7}{*}{\ce{Pd}}& \ce{N2}  & & & &$-0.094$ \\
        & \ce{NH2}  & & & & $-0.453$\\
        & \ce{N2H}  & & & & $1.181$\\
        & \ce{CN}  & & & & $2.073$ \\
        & \ce{NO}  & & & & $1.818$ \\
        & \ce{NH3}  & & & & $-0.184$\\
        & \ce{N}  & & & & $0.727$ \\
        \hline
        \multirow{7}{*}{\ce{TiO2}}& \ce{N2} & $-0.344$ & $-0.271$ & $-0.293$ \\
        & \ce{NH2} & $1.209$ & $1.160$ & $1.246$ &$1.477$\\
        & \ce{N2H} & $1.841$ & $1.878$ & $1.879$ &$1.940$\\
        & \ce{CN} & $5.624$ & $6.024$ & $5.743$ & $5.355$\\
        & \ce{NO} & $3.369$ & $3.411$ & $3.341$ \\
        & \ce{NH3} & $-2.117$ & $-2.039$ & $-1.973$ \\
        & \ce{N} & $4.752$ & $4.400$ & $4.572$ \\
        \hline
        \multirow{7}{*}{\ce{MoO3}}& \ce{N2} & $-0.777$ & $-0.555$ & $-0.519$ \\
        & \ce{NH2} & $-1.743$ & $-1.381$ & $-1.409$ \\
        & \ce{N2H} & $1.239$ & $1.370$ & $1.436$ \\
        & \ce{CN} & $2.375$ & $2.362$ & $2.377$ \\
        & \ce{NO} & $1.334$ & $1.456$ & $1.326$ \\
        & \ce{NH3} & $-2.760$ & $-2.629$ & $-2.525$ \\
        & \ce{N} & $1.305$ & $1.331$ & $1.328$ \\
        \hline
        \multirow{7}{*}{\ce{OCUPUY}}& \ce{N2} & $-1.741$ & $-1.195$ & $-1.394$ \\
        & \ce{NH2} & $-2.627$ & $-2.224$ & $-2.239$ \\
        & \ce{N2H} & $-1.601$ & $-0.647$ & $-1.390$ \\
        & \ce{CN} & $-0.347$ & $0.091$ & $-0.059$ \\
        & \ce{NO} & $0.166$ & $0.894$ & $0.243$ \\
        & \ce{NH3} & $-3.350$ & $-3.030$ & $-2.941$ \\
        & \ce{N} & $-0.261$ & $1.037$ & $0.088$ \\
        \hline
        \multirow{7}{*}{\ce{MIL-125}}& \ce{N2} & $-0.406$ & $-0.322$ & $-0.379$ \\
        & \ce{NH2} & $1.273$ & $1.229$ & $1.265$ \\
        & \ce{N2H} & $0.134$ & $-0.063$ & $0.058$ \\
        & \ce{CN} & $3.787$ & $3.838$ & $3.729$ \\
        & \ce{NO} & $3.231$ & $3.269$ & $3.169$ \\
        & \ce{NH3} & $-1.101$ & $-1.026$ & $-0.995$ \\
        & \ce{N} & $4.789$ & $4.414$ & $4.589$ \\
        \hline
    \hline
  \end{tabular}
\end{table}

\begin{table}[p]
\setlength{\tabcolsep}{4pt}
\renewcommand{\arraystretch}{1.3}
  \caption{Error of the calculated adsorption energy without D3 correction compared to the adsorption energy of optRPA.}
  \label{tab:SIadsorption_err}
  \begin{tabular}{l|*{3}{c}|*{3}{c}|*{3}{c}}
    \hline
    \hline
    \multirow{2}{*}{}   & \multicolumn{3}{c|}{Metals} & \multicolumn{3}{c|}{Metal Oxides} & \multicolumn{3}{c}{MOFs} \\
                        & MaxError & MSE& RMSE   & MaxError & MSE& RMSE   & MaxError & MSE& RMSE\\ 
    \hline
            RPBE        &  $-0.758$ & $0.113 $ & $0.451 $&  $1.157 $ & $0.469 $ & $0.574 $&  $1.932 $ & $0.225 $ & $0.808 $\\
            PBE         &  $-1.010$ & $-0.384$ & $0.517 $&  $0.743 $ & $0.251 $ & $0.406 $&  $-1.185$ & $-0.068$ & $0.511 $\\
            SCAN        &  $-1.357$ & $-0.638$ & $0.725 $&  $-0.847$ & $-0.136$ & $0.322 $&  $1.601 $ & $0.284 $ & $0.774 $\\
            rev-vdW-DF2 &  $-1.225$ & $-0.575$ & $0.674 $&  $-0.808$ & $-0.092$ & $0.335 $&  $-1.510$ & $-0.185$ & $0.667 $\\
            BEEF-vdW    &  $-1.029$ & $-0.282$ & $0.447 $&  $-0.857$ & $-0.155$ & $0.378 $&  $1.416 $ & $0.094 $ & $0.697 $\\
            PBE0        &  $-0.876$ & $-0.088$ & $0.316 $&  $0.825 $ & $0.253 $ & $0.396 $&  $0.886 $ & $0.230 $ & $0.387 $\\
            HSE06       &  $-1.090$ & $-0.132$ & $0.370 $&  $0.655 $ & $0.220 $ & $0.357 $&  $0.846 $ & $0.198 $ & $0.364 $\\
            B3LYP       &  $-0.686$ & $0.078 $ & $0.320 $&  $0.879 $ & $0.439 $ & $0.508 $&  $0.631 $ & $0.284 $ & $0.353 $\\
            RPA@PBE0    &  $-0.277$ & $-0.025$ & $0.143 $&  $0.334 $ & $0.087 $ & $0.157 $&  $0.408 $ & $0.128 $ & $0.233 $\\
    \hline
    \hline
  \end{tabular}
\end{table}

\begin{table}[p]
\setlength{\tabcolsep}{10pt}
\renewcommand{\arraystretch}{1.3}
  \caption{RMSE of the calculated adsorption energy compared to the adsorption energy of optRPA. Unit in \SI{}{eV}.}
  \label{tab:SIadsorption_err_material}
  \begin{tabular}{l|*{6}{c}}
    \hline
    \hline
                        & \ce{Cu} & \ce{Pd} & \ce{TiO2}& \ce{MoO3}& \ce{OCUPUY}& \ce{MIL-125} \\
    \hline
            RPBE-D3     & $0.951 $& $0.767 $& $0.576 $& $0.568 $& $1.177 $& $0.322 $\\
            PBE-D3      & $0.759 $& $0.734 $& $0.328 $& $0.234 $& $0.810 $& $0.283 $\\
            SCAN-D3     & $0.725 $& $0.933 $& $0.486 $& $0.214 $& $0.918 $& $0.361 $\\
            rev-vdW-DF2 & $0.673 $& $0.676 $& $0.423 $& $0.211 $& $0.907 $& $0.261 $\\
            BEEF-vdW    & $0.543 $& $0.323 $& $0.405 $& $0.349 $& $0.957 $& $0.238 $\\
            PBE0-D3     & $0.450 $& $0.509 $& $0.325 $& $0.266 $& $0.316 $& $0.235 $\\
            HSE06-D3    & $0.512 $& $0.575 $& $0.312 $& $0.281 $& $0.305 $& $0.249 $\\
            B3LYP-D3    & $0.460 $& $0.505 $& $0.194 $& $0.368 $& $0.222 $& $0.169 $\\
            RPA@PBE0    & $0.143 $& $      $& $0.102 $& $0.198 $& $0.315 $& $0.096 $\\
    \hline
            RPBE        & $0.460 $& $0.443 $& $0.396 $& $0.708 $& $1.131 $& $0.159 $\\
            PBE         & $0.549 $& $0.484 $& $0.229 $& $0.527 $& $0.694 $& $0.200 $\\
            SCAN        & $0.627 $& $0.811 $& $0.404 $& $0.208 $& $1.054 $& $0.297 $\\
            rev-vdW-DF2 & $0.673 $& $0.676 $& $0.423 $& $0.211 $& $0.907 $& $0.261 $\\
            BEEF-vdW    & $0.543 $& $0.323 $& $0.405 $& $0.349 $& $0.957 $& $0.238 $\\
            PBE0        & $0.229 $& $0.384 $& $0.316 $& $0.462 $& $0.521 $& $0.167 $\\
            HSE06       & $0.281 $& $0.441 $& $0.276 $& $0.423 $& $0.484 $& $0.173 $\\
            B3LYP       & $0.163 $& $0.422 $& $0.240 $& $0.677 $& $0.481 $& $0.136 $\\
            RPA@PBE0    & $0.143 $& $      $& $0.102 $& $0.198 $& $0.315 $& $0.096 $\\
    \hline
    \hline
  \end{tabular}
\end{table}

\begin{table}[p]
\setlength{\tabcolsep}{10pt}
\renewcommand{\arraystretch}{1.3}
  \caption{Standard deviation and error of the calculated adsorption energy without D3 correction compared to the adsorption energy of optRPA, for each material. ``std. dev'' stands for standard deviation between functionals (except for RPA results).}
  \label{tab:SIadsorption_stddev_err}
  \begin{tabular}{l|*{3}{c}}
    \hline
    \hline
                        & std. dev & MSE& RMSE \\ 
    \hline
            \ce{Cu}         & $0.253$ & $-0.301$ & $0.476$\\
            \ce{Pd}         & $0.376$ & $-0.176$ & $0.521$\\
            \ce{TiO2}       & $0.307$ & $0.003$ & $0.345$\\
            \ce{MoO3}       & $0.336$ & $0.310$ & $0.480$\\
            \ce{OCUPUY}     & $0.615$ & $0.322$ & $0.818$\\
            \ce{MIL-125}    & $0.166$ & $-0.056$ & $0.210$\\
    \hline
            Total           & $0.369$ & $0.017$ & $0.510$ \\
    \hline
    \hline
  \end{tabular}
\end{table}

\begin{table}[p]
\setlength{\tabcolsep}{5pt}
\renewcommand{\arraystretch}{1.2}
    \caption{Deviation of calculated adsorption energies from experimental values (in eV), with and without D3 correction.}
    \begin{tabular}{l|*{2}{c}|*{2}{c}}
        \hline
        \hline
        \multirow{2}{*}{}   & \multicolumn{2}{c|}{Cu(100)$+$\ce{NH3}*} & \multicolumn{2}{c}{Pd(111)$+$\ce{NO}*}\\
        \cline{2-5}
                            &  $\Delta_f^{'}E_{ads}^{+D3}-\Delta_f^{'}E_{ads}^{exp}$ & $\Delta_f^{'}E_{ads}-\Delta_f^{'}E_{ads}^{exp}$ &   $\Delta_f^{'}E_{ads}^{+D3}-\Delta_f^{'}E_{ads}^{exp}$ & $\Delta_f^{'}E_{ads}-\Delta_f^{'}E_{ads}^{exp}$\\
         \hline
         RPBE&  $-0.245$ & $0.670$& $-0.788$ & $0.114$\\
         PBE&  $-0.313$ &$-0.014$& $-0.786$ &$-0.504$\\
         SCAN&  $-0.307$ &$-0.172$& $-0.878$ &$-0.756$\\
         rev-vdW-DF2&  \multicolumn{2}{c|}{$-0.052$}& \multicolumn{2}{c}{$-0.866$}\\
         BEEF-vdW&  \multicolumn{2}{c|}{$0.318$}& \multicolumn{2}{c}{$-0.409$}\\
         PBE0&  $-0.361$& $-0.061$& $-0.070$& $0.217$\\
         HSE06&  $-0.390$ & $-0.088$& $-0.044$ &$0.248$\\
         B3LYP&  $-0.137$ & $0.324$& $-0.001$ &$0.464$\\
         RPA@PBE0& \multicolumn{2}{c|}{$0.327$}& \multicolumn{2}{c}{-}  \\
         optRPA& \multicolumn{2}{c|}{$0.117$} &\multicolumn{2}{c}{$0.049$}\\
         \hline
         \hline
    \end{tabular}
    \label{tab:exp_comparison}
\end{table}


\begin{table}[p]
\setlength{\tabcolsep}{6pt}
\renewcommand{\arraystretch}{1.2}
\caption{Comparison of adsorption energy and interaction energy (same gas and solid geometry from the RPBE optimized gas+solid system, and compute single-point calculation to get energies), and the amount of D3 correction added.}
    \begin{tabular}{c|c|cc} 
    \hline
    \hline
         &  property&  single-point& geometry optimization\\ 
         \hline 
         \multirow{2}{*}{\ce{MoO3}+\ce{CN}}&  $\Delta_f^{'}E_{ads}$ (eV)&  $1.06$& $1.45$\\
         &  D3 correction (eV)&  $-0.57$& $-1.46$\\
         \hline
         \multirow{2}{*}{\ce{Pd}(111)+\ce{N2}}&  $\Delta_f^{'}E_{ads}$ (eV)&  $-0.34$& $-0.32$\\
         &  D3 correction (eV)&  $-0.91$& $-0.88$\\
         \hline
         \multirow{2}{*}{\ce{Pd}(111)+\ce{CN}}&  $\Delta_f^{'}E_{ads}$ (eV)&  $1.42$& $1.45$\\
         &  D3 correction (eV)&  $-1.12$& $-1.02$\\
         \hline 
         \hline 
    \end{tabular}
    \label{tab:d3_interaction}
\end{table}

\begin{figure}[p]
    \centering
    \includegraphics[width=\textwidth]{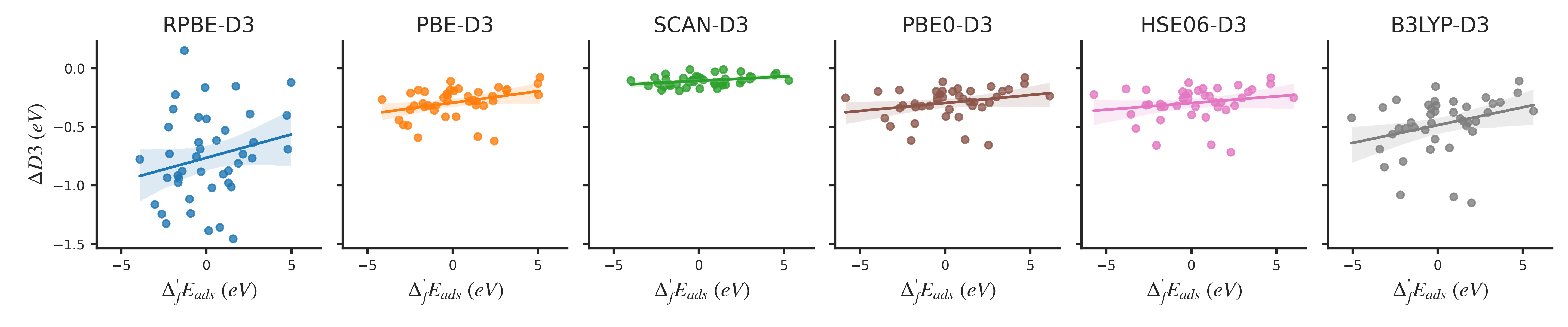}
    \caption{Relation between adsorption energy and D3 correction with regression lines drawn. Confidence interval of 95\% is shaded in color.}
    \label{fig:SIvdw}
\end{figure}

\begin{figure}[p]
    \centering
    \includegraphics[width=\textwidth]{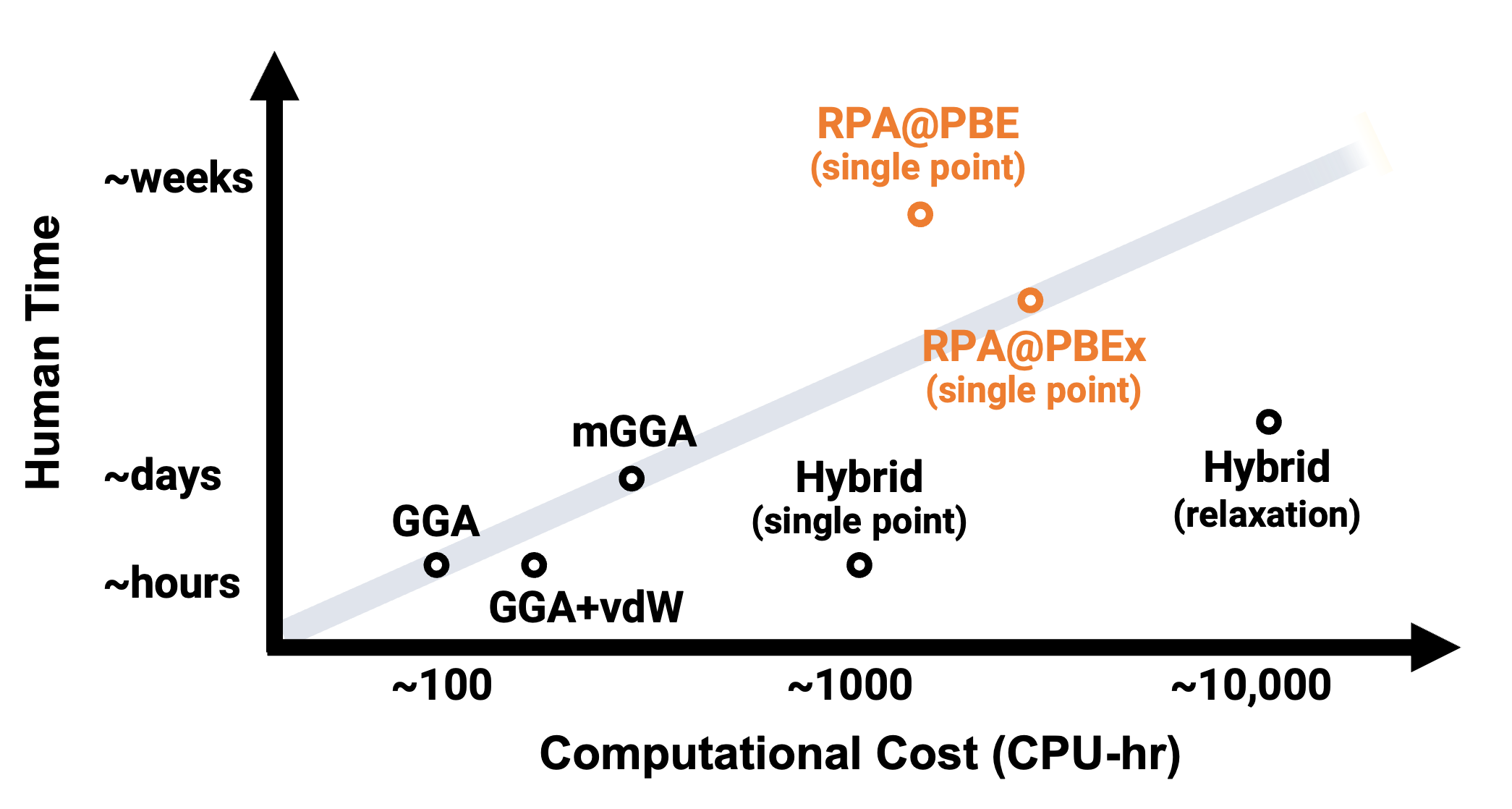}
    \caption{Schematic illustration of computational time and human time required for computing a single adsorption energy for a typical slab system. Human time reflects effort required for convergence testing and post-processing. Representative order-of-magnitude numbers are provided based on our experience in this work. The high memory requirement of RPA is not reflected.
}
    \label{fig:SIcost}
\end{figure}

\begin{table}[p]
\setlength{\tabcolsep}{10pt}
\renewcommand{\arraystretch}{1.3}
  \caption{Multiple adsorption energies (physisorption and chemisorption) of \ce{NO} on \ce{MoO3}(100) compared to that of optRPA chemisorption. Unit in \SI{}{eV}.}
  \label{tab:SIMoO3_NO_physi_chemi}
  \begin{tabular}{l|*{2}{c}}
    \hline
    \hline
                            & chemisorption & physisorption \\ 
    \hline
            RPBE-D3         & $0.384$ & $1.697$ \\
            PBE-D3         & $0.172$ & $1.355$ \\
            PBE0-D3       & $0.146$ & $1.666$ \\
            HSE06-D3       & $0.184$ & $1.594$ \\
            B3LYP-D3     & $0.354$ & $1.273$ \\
    \hline
    \hline
  \end{tabular}
\end{table}

\begin{table}[p]
\setlength{\tabcolsep}{2pt}
\renewcommand{\arraystretch}{1.2}
  \caption{Results of different spin state configurations tested and their energy in OCUPUY, using PBE-D3. Numbers in `Initial magnetic moments' and `Optimized magnetic moments' represent magnetic moment of four vanadium atoms in the unit cell. $\bar{3}$ represents $-3$.}
  \label{tab:optimal_spin_config_OCUPUY}
  \begin{tabular}{l|*{1}{c}|*{1}{c}|*{4}{c}|*{3}{c}}
    \hline
    \hline
    combination of spin states   & \multicolumn{1}{c|}{lowspin} & \multicolumn{1}{c|}{4 up} & \multicolumn{4}{c|}{3 up, 1 down}& \multicolumn{3}{c}{2 up, 2 down}  \\
    \hline
    Initial magnetic moments       & $0000$ & $3333$ & $333\bar{3}$& $33\bar{3}3$  & $3\bar{3}33$& $\bar{3}333$ & $33\bar{3}\bar{3}$& $3\bar{3}3\bar{3}$  & $\bar{3}33\bar{3}$ \\
    Optimized magnetic moments     & $3333$ & $3333$ & $333\bar{3}$& $33\bar{3}3$  & $3\bar{3}33$& $\bar{3}333$ & $33\bar{1}\bar{3}$ & $3\bar{3}3\bar{3}$   & $333\bar{3}$\\
    \hline
    Energy difference from `3333' (eV) & $0.000$& $0.000$ & $0.053$ & $0.052$ & $0.053$ & $0.052$ & $0.716$ & $0.037$ & $0.053$\\
    \hline
    \hline
  \end{tabular}
\end{table}

\begin{table}[p]
\setlength{\tabcolsep}{2pt}
\renewcommand{\arraystretch}{1.2}
  \caption{Different spin states and energy after gas adsorbed in OCUPUY, using PBE-D3. Original state represents the result of Figure 4. `3333' state represents the state where all four vanadium atoms in OCUPUY unitcell have up-spin states. $\Delta$E stands for the energy difference of the `3333' state compared to the original state (unit in eV). `mag' stands for the total magnetic moment. Since \ce{NH2} and \ce{CN} are already in the `3333' state (or $\bar{3}\bar{3}\bar{3}\bar{3}$), the last row of them is empty.}
  \label{tab:OCUPUY_gas_mag}
  \begin{tabular}{l|*{2}{c}|*{2}{c}|*{2}{c}|*{2}{c}|*{2}{c}|*{2}{c}|*{2}{c}}
    \hline
    \hline
    \multirow{2}{*}{}   & \multicolumn{2}{c|}{\ce{N2}*} & \multicolumn{2}{c|}{\ce{NH2}*} & \multicolumn{2}{c|}{\ce{N2H}*} & \multicolumn{2}{c|}{\ce{CN}*} & \multicolumn{2}{c|}{\ce{NO}*} & \multicolumn{2}{c|}{\ce{NH3}*} & \multicolumn{2}{c}{\ce{N}*} \\
    \cline{2-15}
                     & $\Delta$E  & mag & $\Delta$E   & mag& $\Delta$E   & mag& $\Delta$E   & mag& $\Delta$E   & mag& $\Delta$E   & mag& $\Delta$E   & mag\\ 
    \hline
    Original state   & $0.0$   & $5.7$ & $0.0$ & $-11$& $0.0$  & $3$ & $0.0$& $11$& $0.0$& $2.4$& $0.0$ & $6.5$& $0.0$ & $-0.9$ \\
    `3333' state     & $0.210$ & $11.8$ & -     & -  & $0.113$& $9.0$ & -  & -  & $-0.020$ & $9.0$ & $-0.001$ & $12.0$ & $-0.067$ & $9.0$\\
    \hline
    \hline
  \end{tabular}
\end{table}

\end{document}